\documentclass[preprint2]{aastex}

\newcommand{\zabs}{$z_{\rm abs}\,$}

\newcommand{\kms}{km~s$^{-1}\,$}
\newcommand{\cm}{cm$^{-2}\,$}
\newcommand{\cmm}{cm$^{-3}\,$}

\slugcomment{Submitted to ApJ}

\shorttitle{Metal absorbers toward Q0347--3819 and 
APM BR J0307--4945}
\shortauthors{Levshakov et al.}

\def\la{\;
\raise0.3ex\hbox{$<$\kern-0.75em\raise-1.1ex\hbox{$\sim$}}\; }
\def\ga{\;
\raise0.3ex\hbox{$>$\kern-0.75em\raise-1.1ex\hbox{$\sim$}}\; }

\def\la{\;
\raise0.3ex\hbox{$<$\kern-0.75em\raise-1.1ex\hbox{$\sim$}}\; }
\def\ga{\;
\raise0.3ex\hbox{$>$\kern-0.75em\raise-1.1ex\hbox{$\sim$}}\; }

\begin{document}
\title{Metal abundances and kinematics of quasar absorbers --
II. Absorption systems
toward Q0347--3819 and APM BR J0307--4945\altaffilmark{1}
}

\author{S. A. Levshakov\altaffilmark{2,3},
I. I. Agafonova\altaffilmark{3}, 
S. D'Odorico\altaffilmark{4},
A. M. Wolfe\altaffilmark{5},
}
\and
\author{M. Dessauges-Zavadsky\altaffilmark{4,6}
}
\altaffiltext{1}
{Based on observations obtained at
the VLT Kueyen telescope (ESO, Paranal,
Chile) and at the W. M. Keck Observatory  (jointly operated
by the University of California, the California Institute of
Technology and the National Aeronautics and Space Administration).
}
\altaffiltext{2}{Division of Theoretical Astrophysics,
National Astronomical Observatory, Tokyo 181-8588, Japan}
\altaffiltext{3}{Department of Theoretical Astrophysics,
Ioffe Physico-Technical Institute, St. Petersburg 194021, Russia}
\altaffiltext{4}{European Southern Observatory, 
Karl-Schwarzschild-Str. 2,
D-85748 Garching bei M\"unchen, Germany}
\altaffiltext{5}{Department of Physics, Code 0354, 
and Center for Astrophysics and Space Sciences,
Code 0424, University of California at San Diego, 9500 Gilman Drive,
La Jolla, CA 92093
}
\altaffiltext{6}{Observatoire de Gen\`eve, CH-1290, Sauverny,
Switzerland
}

\begin{abstract}

Detailed Monte Carlo Inversion  
analysis of the spectral lines from three Lyman limit systems (LLS)\,
[$N$(H\,{\sc i}) $\ga 1.0\times10^{17}$ \cm] and nine lower $N$(H\,{\sc i})
systems [$2\times10^{14}$ \cm $\la N$({\rm H}\,{\sc i}) $\la
2\times10^{16}$ \cm]  
observed in the VLT/UVES spectra of Q0347--3819 (in the range
$2.21 \leq z \leq 3.14$) and of APM BR J0307--4945 
(at $z = 4.21$ and 4.81) is presented.
Combined with the results from a previous work,     
the analyzed LLSs show that they are a {\it heterogeneous} population
originating in different environments.
A functional dependence of the line-of-sight velocity dispersion
$\sigma_{\rm v}$ on the absorber size $L$ is confirmed: the
majority of the analyzed systems follow the scaling relation 
$\sigma_{\rm v} \sim (N_{\rm H}\,L)^{0.3}$ (with $N_{\rm H}$ being
the total gas column density). This means that most absorbers
may be related to virialized systems like galaxies or their halos.  
Previously noted enhancement of the metal content in small size systems is
also confirmed: metallicities of $Z \sim (1/3-1/2)\,Z_\odot$\, are found in systems
with $L \la 0.4$ kpc, whereas we observe much lower metal abundances in systems
with larger linear sizes.
For the first time in LLSs, a pronounced
[$\alpha$-element/iron-peak] enrichment is revealed: 
the absorber at \zabs = 2.21 shows
[O/Fe]  = $0.65\pm0.11$,
[Si/Fe] = $0.51\pm0.11$, and [Mg/Fe] = $0.38\pm0.11$.
Several absorption systems exhibit characteristics 
which are very similar to that observed in high-velocity 
clouds in the Milky Way and may be considered as high-redshift counterparts 
of Galactic HVCs.
\end{abstract}


\section{Introduction}

With the present  paper we continue to study  the chemical composition and 
the kinematic characteristics of quasar
absorption systems using a new computational procedure, --
the Monte Carlo Inversion algorithm (MCI), -- developed earlier in a series of papers 
[see Levshakov, Agafonova \& Kegel (2000); hereafter LAK].
The MCI technique allows us 
to recover self-consistently the physical parameters 
of the intervening gas cloud (such as the average gas number density $n_0$,
the column densities for different species $N_{\rm a}$, the kinetic temperature
$T_{\rm kin}$, the metal abundances $Z_{\rm a}$, and the linear size $L$),  
the statistical characteristics of the underlying 
hydrodynamical fields (such as the line-of-sight velocity dispersion
$\sigma_{\rm v}$, and the density dispersion $\sigma_{\rm y}$), and
the line of sight density $n_{\rm H}(x)$ and
velocity $v(x)$ distributions (here $x$ is the dimensionless 
coordinate in units of $L$).
Having this comprehensive 
information we are able to classify the absorbers more reliably and 
hence to obtain important clues concerning the physical conditions in 
intervening galaxies, galactic 
halos and large scale structure objects at high redshifts. 
Besides, it will also be possible to constrain 
the existing theories of the origin of  
galaxy formation
since the observed statistics of the damped Ly$\alpha$ (DLA) and Lyman limit (LLS)
systems is believed to be a strong test of different cosmological models
(e.g. Gardner et al. 2001; Prochaska \& Wolfe 2001).

In the first part of our study (Levshakov et al. 2002a, hereafter Paper I)
we reported results
on the absorption systems at \zabs = 1.87, 1.92 and 1.94  toward the
HDF-South quasar J2233--606. These
systems  exhibit many metal lines with quite complex structures.
It was found that 
all profiles can be well described with an assumption of a homogeneous
metallicity and a unique photoionizing background.
According to the estimated sizes, velocity dispersions
and  metal contents 
the absorbers at \zabs = 1.92 and 1.87 were related to the galactic
halos whereas the system at \zabs = 1.94 was formed, more likely, in an
irregular star-forming galaxy.
It was also found, 
that the linear size and the line-of-sight velocity dispersion for all
three absorbers obey a scaling relation of the same kind that can be expected
for virialized systems.

The present paper deals 
with absorbers observed in the spectra of  Q0347--3819 ($z_{\rm em} = 3.23$)
and  APM BR J0307--4945 ($z_{\rm em} = 4.75$, see \S~2.1).
Both spectra include several dozens of 
systems containing metals, but most of them are
weak and severely blended and hence do not allow to estimate the underlying physical 
parameters with a reasonable accuracy.  
After preliminary analysis only 
12 systems were chosen for the inversion with the MCI
and their properties are described below.

The structure of the paper is as follows. \S~2 describes the data sets.  
In \S~3 our model assumptions and basic equations are specified.
The estimated parameters for individual systems are given in
\S~4. The implication of the obtained results 
to the theories of LLS origin are discussed in \S~5 and  
our conclusions are reported in \S~6.
Appendix contains a table with typical parameters of different absorbers
which are referred to in the present study.

\section{Observations}

The spectroscopic observations of Q0347--3819 and APM BR J0307--4945  obtained 
with the UV-Visual Echelle Spectrograph 
UVES (D'Odorico et al. 2000) on the VLT/Kueyen 8.2 m telescope
are described in detail
by D'Odorico, Dessauges-Zavadsky \& Molaro (2001) and by
Dessauges-Zavadsky et al. (2001), respectively.
Both spectra were observed with the spectral resolution
FWHM $\simeq 7$ \kms.
For the analysis of metal systems from the Q0347--3819 spectrum with lines in the
range 4880 -- 6730 \AA\, which was not covered by the VLT observations,
we used a portion of the Q0347--3819 spectrum obtained with the 
High-Resolution Echelle Spectrograph
HIRES (Vogt et al. 1994) 
on the 10~m W. M. Keck~I telescope (Prochaska \& Wolfe 1999). The spectral
resolution in this case was about 8 \kms. 
The VLT/UVES data are now available for public use in the VLT data archive.

The majority of the metal systems in the spectrum of Q0347--3819 were identified in
Levshakov et al. (2002b), whereas the \zabs = 4.21 system toward 
APM BR J0307--4945 was distinguished by 
Dessauges-Zavadsky et al.\footnote{Data are listed in Table~5 which is only
available in electronic form at the CDS via anonymous ftp to
cdsarc.u-strasbg.fr (130.79.128.5).} 
(2001) as consisting of two sub-systems~: one at \zabs = 4.211 and the other at
\zabs = 4.218. A new system at \zabs = 4.81 is analyzed here for the first time.

\subsection{Emission redshift of APM BR J0307--4945}

The emission redshift of this distant quasar $z_{\rm em} = 4.728\pm0.015$ 
was previously
measured by P\'eroux et al. (2001) from the 
\ion{Si}{4}+\ion{O}{4}] $\lambda1400.0$ and \ion{C}{4} $\lambda1549.1$ 
lines observed in
the $\sim 5$ \AA\, resolution spectrum obtained with the 4~m Cerro 
Tololo Inter-American Observatory telescope. 

In our VLT/UVES spectrum of this quasar 
a few additional lines can be identified which are
useful for the redshift measurements. 
The most important of them is the weak \ion{O}{1}
$\lambda1304$ line. From earlier studies 
(see, e.g., Tytler \& Fan 1992 and references cited
therein) it is known that `low-ionization lines' 
such as \ion{O}{1} $\lambda1304$ are systematically
redshifted and narrower than `high-ionization' lines 
such as \ion{C}{4}, \ion{Si}{4}, and Ly$\alpha$.

In Fig.~1 we compare the \ion{O}{1} profile 
with those of \ion{C}{4} and of a wide blend
of the Ly$\alpha$+\ion{N}{5}+\ion{Si}{2} lines. 
All these lines are shown in the same velocity
scale which is defined by the \ion{O}{1} $\lambda1304$ center corresponding to
$z_{\rm em} = 4.7525$. 
This \ion{O}{1} line is redshifted 
with respect to the $z_{\rm em}$ value deduced by P\'eroux et al.
from the measurements of the \ion{Si}{4}+\ion{O}{4}] and \ion{C}{4} profiles.
Because the Ly$\alpha$ emission line is blended with other emission
lines and its blue wing is distorted 
by numerous narrow absorption lines, profile comparison
cannot be very accurate in this case. 
Nevertheless, we found a smooth fit to the \ion{C}{4}
$\lambda1549$ profile (which is unblended and shows 
significant asymmetry) and used this
synthetic profile for comparison with other lines 
(altering the amplitude of the synthetic
profile to match the line profile while keeping its center unchanged).

Fig.~1 shows that this simplified procedure 
indeed allows us to achieve a fairly good concordance between 
the \ion{C}{4} line and the  Ly$\alpha$+\ion{N}{5}+\ion{Si}{2} blend.
This could indicate that the redshift of the quasar is higher than that
measured by P\'eroux et al., 
being actually $z_{\rm em} \simeq 4.753$.

\section{Model assumptions and the MCI procedure}

The complete description of the MCI code is given in LAK and
its most updated version -- in Paper~I. Since this technique is
relatively new, we briefly outline it here and stress its difference
from the Voigt profile fitting (VPF) procedure
commonly used for the QSO absorption line analysis.

The VPF deconvolution is based on the 
assumption that the observed complex line profiles
are caused by several separate clouds randomly distributed along the line
of sight. In every cloud, gas is characterized by the constant density and
normally distributed velocities (the $b$ parameter usually estimated in the
VPF procedure just stands for the dispersion of the velocity distribution).
Because of 
the constant gas density, the ionizing structure inside the cloud is described
by a single ionization parameter $U$ which can be estimated from the measured
column densities of lines of different ions with help of some photoionization
code if the spectrum of the background ionizing radiation is given.

However, numerous cosmological hydrodynamical calculations performed in the
previous decade have shown that the QSO absorption lines arise more likely 
in the smoothly fluctuating intergalactic medium 
in a network of sheets, filaments, and
halos (e.g., Cen et al. 1994; Miralda-Escud\'e et al. 1996;
Theuns et al. 1998). 
This model finds its support also 
in modern high resolution spectroscopic observations:
the increasing spectral resolution reveals progressively 
more and more complex profiles. 
A very important characteristic of the continuous absorbing medium is that
the contribution to any point within the line profile comes
not only from a single separate area (a `cloud') 
but from all volume elements (`clouds')
distributed along the sightline within the absorbing region and having the same
radial velocity
(see, for details, \S~2.2 in LAK). 

If the absorption systems are indeed formed in the fluctuating IGM, then 
the above described VPF procedure which interprets
each absorption feature in the line profile as caused by one distinguished cloud
is not, in general, physically justified.
In some systems, this approach can produce rather erratic
results like extremely varying metallicities between subcomponents, negative
kinetic temperatures, exotic UV background spectra etc. 
(see examples in Levshakov et al. 1999; LAK; Paper~I).

The MCI procedure is based on the assumption that  
all lines observed in a metal system arise
in a continuous absorbing gas slab of a thickness $L$ with
a fluctuating gas density and a random velocity field.  
We also assume that
within the absorber the metal abundances are constant,
the gas is optically thin for the ionizing UV radiation, and
the gas is in the thermodynamic and ionization equilibrium.
The last assumption means that
the fractional ionizations of different ions are determined
exclusively by the gas density  and vary from point to point
along the sightline.
These fractional ionization variations are just the cause
of the observed diversity of profile shapes between ions of low- and
high-ionization stages.

Whereas most of the above mentioned assumptions are quite natural, 
that of the constant
metallicity over the entire absorbing region needs additional
discussion. On one hand, it is required from mathematical point of view.
Namely, the splitting of the velocity and density fields is effective
if all observed ions share the same velocity distribution but respond
differently to the gas density. If in addition to the varying density and velocity 
one would allow for the varying metallicity, 
the inverse problem becomes fully degenerate, 
i.e. it would have infinitely large number of solutions. 
On the other hand, the constant metallicity has some observational support:
results obtained in numerous studies
of the Galactic halo chemical composition reveal no
systematic differences in the gas-phase abundances within 
galactocentric distances of $7-10$ kpc in various directions
(e.g., Savage \& Sembach 1996).
But, of course, we cannot exclude the case when the line of sight passes
through many types of environments 
with different enrichment histories within a given
absorber. If the metallicities within such an absorber differ only slightly
($\la 0.5$ dex), the observed lines of different ions can be well fitted to the
synthetic profiles calculated with some average value of the metal content.
If, however, the differences in the metallicities are really large ($\ga 1$ dex),
the self-consistent fitting of all observed lines becomes impossible.
In this case we have to split the absorber into separate regions having 
different metal abundances (see \S~4.2.2 and 4.3.1 for examples).

It is well known that the measured metallicities depend in a crucial way 
on the adopted background ionizing spectrum. We started in all cases with
the Haardt-Madau (HM) background ionizing spectra (Haardt \& Madau 1996)
computing the fractional ionizations and the kinetic temperatures 
with the photoionization code CLOUDY (Ferland 1997). 
If the fitting with the HM spectrum was impossible, we used another spectra, e.g.
the Mathews and Ferland (MF) spectrum (Mathews \& Ferland 1987). 

The MCI procedure itself is implemented in the following way.
Within the absorbing region the radial velocity  $v(x)$ and
the total hydrogen density $n_{\rm H}(x)$ 
along the line of sight are considered as two random fields 
which are represented by their sampled values
at equally spaced intervals $\Delta x$,
i.e. by the vectors
$\{ v_1, \ldots, v_k \}$ and $\{ n_1, \ldots, n_k \}$
with $k$ large enough ($\sim 150-200$)
to describe the narrowest components of the complex spectral lines.
The radial velocity is assumed to be normally distributed with the
dispersion $\sigma_{\rm v}$, whereas the gas density is distributed
log-normally with the mean $n_0$ 
and the second central dimensionless moment $\sigma_{\rm y}$
($y = n_{\rm H}/n_0$). 
Both stochastic fields are calculated using the Markovian processes (see LAK for
mathematical basics).
The model parameters estimated 
in the least-squares minimization  of the objective function
(see eqs.[29] and [30] in LAK) include
$\sigma_{\rm v}$ and $\sigma_{\rm y}$ along with
the total hydrogen column density $N_{\rm H}$,
the mean ionization parameter $U_0$,
and the metal abundances $Z_a$ 
for $a$ elements observed in a given absorption-line system.

The computations are carried out in two steps: firstly a point
in the parameter space ($N_{\rm H}, U_0, \sigma_{\rm v}, \sigma_{\rm y},
Z_{\rm a})$  is chosen at random and then an optimal configuration
of $\{v_i\}$ and $\{n_i\}$ for this parameter set is searched for.
These steps are repeated till a minimum of the objective function 
($\chi^2 \sim 1$ per degree of freedom) is achieved. 
To optimize the configurations of  $\{v_i\}$ and $\{n_i\}$,
the simulated annealing algorithm with Tsallis acceptance
rule (Xiang et al., 1997) and an adaptive annealing temperature choice
is used (details are given in Paper~I).

The following important fact should be taken into account when
one interprets the results obtained with the MCI technique.
The mean ionization parameter $U_0$ is related to the parameters of the
gas cloud as (see eq.[28] in LAK)
\begin{equation}
U_0 = \frac{n_{\rm ph}}{n_0} (1 + \sigma^2_{\rm y})\; .
\label{eq:F1}
\end{equation}
Here $n_{\rm ph}$ is the number density of photons with energies above 1 Ry which is
determined by the intensity of the adopted background ionizing spectrum.

This equation shows that if the density field is fluctuating ($\sigma_{\rm y} > 0$),
then with the same mean density $n_0$ and the same background ionizing spectrum
we obtain a higher value of $U_0$ without any additional sources of the ionization.
Intermittent regions of low and high ionization caused by the density fluctuations
will occur in this case along the sightline. On the other hand, for a given $U_0$
the mean gas density $n_0$  is also higher in the fluctuating media as compared to
the completely homogeneous gas clouds ($\sigma_{\rm y}$ = 0). 
Since the linear size of the
absorber is $L = N_{\rm H}/n_0$, the sizes estimated  with the   
assumption of a constant density (as, e.g., in the VPF) may occur too large. 

Another important question is whether the MCI solution is unique and accurate.
In general, the inverse problems are highly non-linear and ill-posed which implies
multiple solutions and/or very broad uncertainty ranges 
for the recovered parameters. 
To produce a physically reasonable solution, one has to account for all available
information related to the case under study. For instance, the more lines of
different ionic transitions are included into the analysis, the more accurate
result can be obtained since both 
low- and high-density regions are probed simultaneously.
One may also compare the relative metal abundances predicted 
by nucleosynthetic theories
with those provided by the MCI. An odd pattern may indicate a misleading solution.
The recovered linear sizes must also be in agreement 
with the characteristic sizes of the
absorbers stemming from observations 
of gravitationally lensed quasars and quasar pairs
which show $L \la 100$ kpc. 

One of the main problem in the analysis of the QSO high redshift spectra
is the line blending which hampers significantly the inversion of the
observed spectra.
As compared to the VPF method,
the MCI is much more robust dealing with blended lines
due to the assumption that all ions trace the same underlying gas density
and velocity distributions.
This means that we are able to reconstruct both
distributions using unblended parts of different lines.
It is obvious that the accuracy of the recovered
parameters improves with increasing number of lines and the variety of
ions involved in the analysis.
A priori we do not know which parts of the lines are blended and which
are not. To clarify this, several test runs with different
arrangements of lines are carried out till a self-consistent fit
for the majority of lines observed in the spectrum is found.

\section{Results on individual metal systems}

All results given below in Tables~1-3 were obtained using the MCI procedure as 
described in Paper~I. 
Giving the shape and the intensity at 1 Ry of the background ionizing radiation,
the errors of the fitting parameters $U_0$, $N_{\rm H}$, $\sigma_{\rm  v}$, 
$\sigma_{\rm y}$ 
and $Z_{\rm a}$  are about 15-20\% , the errors of the
estimated column densities are less than 10\%, whereas 
the derived parameters $n_0$ and $L$ are estimated with
about 50\% accuracy.
These errors, however, should be considered as internal in 
the sense that they reflect merely the configuration 
of the parameter space in the vicinity of a minimum of
the objective function. 
To what extent the recovered parameters may correspond to their real values
is discussed separately for each individual absorbing system.
We note in passing 
that the density $n_0$ and by this the linear size $L$ scales
with the intensity of the radiation field (see eq.[1]).

The analyzed metal systems are described within three categories~:
(1) Lyman limit systems with 
$N$(\ion{H}{1}) $> 5\times10^{16}$ \cm, 
(2) Ly$\alpha$ absorbers with $N$(\ion{H}{1}) $< 5\times10^{16}$ \cm,\, and
(3) Ly$\alpha$ systems with a probable metallicity gradient.
Their physical properties are compared with different types of absorbers
listed in Appendix.

\subsection{Lyman limit systems}

\subsubsection{Q0347--3819, \zabs = 2.21}

This system consists of a broad saturated Ly$\alpha$ hydrogen 
line spread over 500 \kms\,  and of metal lines of low and high
 ionized species~:  \ion{O}{1}\,$\lambda1302$, \ion{C}{2}\,$\lambda1334$, 
\ion{Mg}{2}\,$\lambda\lambda2796,$ 2803, \ion{Al}{2}\,$\lambda1670$, 
\ion{Si}{2}\,$\lambda\lambda1190,$ 1193, 
\ion{Fe}{2}\,$\lambda\lambda1608,$ 2344, 2382, 2586, 2600, 
\ion{Al}{3}\,$\lambda1854$, \ion{Si}{3}\,$\lambda1206$ and
\ion{C}{4} and \ion{Si}{4} doublets as well.
The physical parameters obtained with the MCI are presented in Table~1
whereas the corresponding observed and synthetic spectra are shown in Fig.~2. 
The recovered density and velocity distributions
along the line of sight are plotted in Fig.~3. The intermittent high- and
low-density regions giving rise to, respectively, low- and high-ionization
species (a multiphase medium) are clearly seen. 
This means that the lines of different ionization stages arise in 
different areas despite of having the same radial velocities [see, e.g., the
regions with $v(x) \simeq 0$ \kms\, in Fig.~3].

According to these data, the system under
study is a compact (380 pc) warm cloud ($T_{\rm kin}  \simeq 9000$ K) with a 
high metal content (0.6 solar value for O) and a high velocity dispersion 
($\sigma_{\rm v} \simeq 80$ \kms). 

The inferred \ion{H}{1} column density of 
$4.6\times10^{17}$ \cm\, classifies this system as a typical 
LLS [since $\log N$(\ion{H}{1}) $> 17$ \cm]. In principle, this value lies beyond 
the applicability limit of the MCI which is formally valid for 
$\tau_{912} < 1$\, [if $N$(\ion{H}{1}) = $4.6\times10^{17}$ \cm\,, then  
$\tau_{912} = 3$]. 
Besides, having only one saturated Ly$\alpha$ line we cannot in any case  
say for sure that the estimated value is the real hydrogen column density. 
But there are also reasons that  make the
obtained solution quite plausible. 

First, we consider an absorber as a clumpy region which implies 
that the ionizing 
radiation may penetrate the cloud from different directions without 
being significantly altered 
(i.e. we assume that the density of the background ionizing
radiation is not reduced much and its spectral distribution is not changed 
considerably in a gas cloud which is not a slab of a uniform density). 
Second, the observed mixture of the
\ion{O}{1} and \ion{Fe}{2} lines and 
the \ion{C}{4} and \ion{Si}{4} lines makes it possible
to fix the mean ionization
parameter $U_0$ quite strictly because the fractional ionization curves 
for, e.g., \ion{O}{1} and \ion{C}{4} are very different. Thus, the \ion{H}{1}
column density can hardly be less than the value estimated by the MCI since 
in that case the metallicity would be higher than solar.
Besides, the same velocity interval covered by
both the Ly$\alpha$ and \ion{Si}{3}\,$\lambda1206$ lines suggests that  
we observe probably the real profile of the \ion{H}{1} Ly$\alpha$ 
(a blend with the Ly$\beta$ line from the system 
at \zabs = 2.8102 which falls in the range $-40$ \kms $< \Delta v < 140$ \kms\, 
has a little influence on the Ly$\alpha$ red wing  -- see Fig.~2). 
Higher values of $N$(\ion{H}{1}) cannot be excluded, but
additional calculations have shown that
the maximum of the \ion{H}{1} column density is limited~: both available wings of
the Ly$\alpha$ 
line do not allow an increase in $N$(\ion{H}{1}) by more than a factor of three.

This uncertainty does not change, fortunately, the main characteristic 
of this system -- we do observe at \zabs = 2.21398 
a compact ($L < 1$ kpc) metal-rich ($Z > 0.1\,Z_\odot$) cloud.
As seen from Fig.~2, the synthetic profiles represent well all unblended spectral 
features (broad absorption features at the position of the \ion{C}{4} doublet are
not consistent with each other and, hence, they cannot be attributed to 
the real \ion{C}{4} profiles; the same is valid for the \ion{Si}{4} doublet).
The measured relative abundances are discussed later in \S~5.3. 
Here we note that
a set of the identified species and 
their pattern as well as the estimated mass ($\sim 10^4 M_\odot$)
resemble parameters 
measured in so-called high and intermediate velocity clouds (HVC, IVC) in the Local 
Group (Wakker 2001). 
Although the nature of the HVCs and IVCs  is still poorly understood,  
they are unlikely to be phenomena isolated to the Milky Way. 
The system at \zabs = 2.21 may be one of similar 
HVCs likely to be encountered in an high-redshift
galactic halo. A more precise determination of its properties would require
the higher Lyman series lines to be included in the analysis
(these lines, however, can be observed with space telescopes only).

\subsubsection{Q0347--3819, \zabs = 2.81}

Here we identified three neutral hydrogen lines and several metal absorption lines
both of low and high ionic transitions (see Fig.~4, the \ion{C}{4} doublet falls,
unfortunately, in a wavelength coverage gap of the Keck spectrum). 
The estimated physical parameters are listed in Table~1 whereas the corresponding
synthetic spectra are shown in Fig.~4 by the smooth curves.
It should be noted, however, that the solution with the 
\ion{H}{1} column density of $5.5\times10^{16}$ \cm\, 
is not unique because all
available hydrogen lines are saturated and partly blended in the red wings.
We found also another solution with $N$(\ion{H}{1}) $= 10^{17}$ \cm.
The solution presented in Table~1 
has been chosen because it delivered a more or less consistent set
of all parameters: the velocity dispersion $\sigma_{\rm v} \simeq 60$ \kms,
the mean gas density $n_0 \simeq 10^{-3}$ \cmm,
the linear size $L \simeq 30$ kpc, and the metallicity $Z \sim 1/10Z_\odot$. 
For comparison, the second solution with $N$(\ion{H}{1}) $=10^{17}$ \cm\, gives
the metallicity $Z \simeq 1/20Z_\odot$, the size $L \simeq 70$ kpc and the same
other parameters.
According to the both sets of the estimated parameters 
and the fact that rather strong low-ionization lines of \ion{C}{2} 
and \ion{Si}{2} are observed,
the \zabs = 2.81 system could be related to an inner galactic halo.

\subsubsection{APM BR J0307--4945, \zabs = 4.21}

In the available spectral range 
we identified hydrogen lines Ly$\alpha$, Ly$\beta$ and 
Ly$\gamma$ as well as metal lines  \ion{C}{2}\,$\lambda1334$, 
\ion{Si}{2}\,$\lambda1526$, 
\ion{C}{3}\,$\lambda977$, \ion{N}{3}\,$\lambda989$, 
\ion{Al}{3}\,$\lambda1854$, \ion{Si}{3}\,$\lambda1206$ and the
\ion{C}{4} and \ion{Si}{4} doublets.
From the metal line profiles, 
two main subsystems can be clearly 
distinguished: the first at $v \simeq -200$ \kms\,,
and the second at $v \simeq 200$ \kms (see Fig.~5). 
The analysis of the \zabs = 4.21 absorber
has been carried out in two steps: ($i$) the subsystem at $v \simeq 200$ \kms\, was
treated separately 
(the results are listed in column 4 in Table~1 and the corresponding
synthetic spectra are shown by the solid lines in Fig.~5),
($ii$) both systems were fitted together 
(column 5 in Table~1, the dotted lines in Fig.~5).

The recovered metallicities are high (about $1/3Z_\odot$ at $v \simeq 200$ \kms\,,
and slightly higher at $v \simeq -200$ \kms) and their pattern is nearly solar. 
At low redshifts similar characteristics are measured, e.g.,  
in high-metallicity blue compact galaxies. 
At high redshifts, some radio galaxies are known which show two spatially 
resolved emitting regions, slightly sub-solar metallicities and sizes of 
about tens of kpc, e.g., MRC 2104--242 at $z = 2.49$ (Overzier et al. 2001) or 
TN J1338--1942 at $z = 4.11$ (De Breuck et al. 1999). 
It is also known that
galaxies showing morphological evidence of a merger have 
excessive velocity widths of their spectral lines (e.g.
Mall\'en-Ornelas et al. 1999). 

Thus we conclude that the LLS under study probably arises 
when the line of sight intersects two 
clumps which may be merging. 
The observed range of the metal lines (from $\simeq -300$ to 400 \kms) and their
very complex structures are
also in line with this picture. 

Although the recovered values are self-consistent, we cannot guarantee their
uniqueness because of the following reasons. The hydrogen lines are blended and
saturated and, hence, the real $N$(\ion{H}{1}) value may be higher leading to
lower metal abundances. Besides, in our calculations the HM background ionizing
spectrum was adopted. However, it is probable that in close pair of galaxies
where a significant star-forming activity 
is triggered by the merging, the HM spectrum
can be affected by the local sources. In particular, the discrepancies between
the observed and theoretical intensities seen
in the \ion{C}{4} and \ion{Si}{4} components at $v = 100$ \kms\, 
as well as the estimated overabundant ratio of [C/Si] $\simeq 0.1$ 
(usually [Si/C] lies between 0 and 0.4, see Appendix)
may be caused by
inadequate choice of the background ionizing spectrum. 
Nevertheless, the observed metal profiles are quite consistent with the HM spectrum  
and therefore we expect that the influence of the local sources may not be very
strong and, hence, the interpretation of this absorption system will not be
significantly altered.

\subsection{Absorbers with $N$(\ion{H}{1}) $< 5\times10^{16}$ \cm}

\subsubsection{Q0347--3819, \zabs = 2.53 and 2.65}

These two metal systems with \zabs = 2.5370 and 2.65044 show
broad and saturated hydrogen Ly$\alpha$ lines 
and lines of silicon and carbon in different ionization stages. 
Computational results are presented in Table~2, the observed and synthetic profiles  
are shown in Figs.~6 and~7. 

The solutions listed in Table~2 are non unique 
because of the sole hydrogen line and blended
low-ionization lines of \ion{Si}{2}\,$\lambda$1260  and \ion{C}{2}\,$\lambda$1334. 
If in reality the \ion{Si}{2}\,$\lambda$1260 absorption 
is absent at \zabs = 2.5370, then
the solution with a higher $U_0$ and, hence, 
with a large linear size can also be obtained. 
Therefore we consider the estimated linear size of 13 kpc as a lower limit.
The system at \zabs = 2.650 may have low-ionization 
\ion{C}{2}\,$\lambda1334$ absorption resulting in a lower mean ionization parameter 
and a smaller linear size as compared with those listed in Table~2.
A rather high overabundance ratio [Si/C] $\simeq 0.5$ estimated for this system 
also allows us to speculate that the real ionization parameter may be lower.
But these considerations do not change the classification of both absorbers:   
most probably they are hosted by halos of some distant galaxies.

\subsubsection{Q0347--3819, \zabs = 2.962 and 2.966}

These two systems are separated by only 300 \kms\, but demonstrate 
very different physical characteristics (see Table~2 and Figs.~8 and 9). 
Multiple hydrogen lines available in their spectra make it possible to
estimate quite accurately the hydrogen column densities and all other physical
parameters. 

The \zabs = 2.96171 system shows a broad Ly$\alpha$ line 
extending over 400 \kms\, and weak absorption lines 
of highly ionized silicon and carbon [the \ion{C}{3}\,$\lambda977$ line
is contaminated by the Ly$\gamma$ line from the \zabs = 2.97915 system (see Fig.~11);
the \ion{C}{2}\,$\lambda$1334 line is 
in a wavelength coverage gap of the Keck spectrum; the \ion{Si}{2}\,$\lambda\lambda$
1260, 1193 lines are strongly blended]. 
The derived physical parameters are typical for a galactic halo absorber: 
a low density ($n_0 \simeq 3\times10^{-4}$ \cmm), 
low metallicity ($Z \simeq  0.01\,Z_\odot$), hot ($T_{\rm kin} \simeq 40000$ K) 
cloud of $L \simeq 20$ kpc size. The large overabundance of silicon as compared with
carbon ([Si/C] $\la 0.8$) can be explained by the uncertainty in the estimated
carbon abundance: only one weak \ion{C}{4}\,$\lambda1548$ line is available for
the analysis.

The adjacent system at \zabs = 2.96591 is on the contrary a very compact 
($L \simeq 120$ pc), warm ($T_{\rm kin} \simeq 10000$ K) cloud, 
5 times more denser and 30 times more metal abundant.
This absorber reveals also a weak 
\ion{N}{3}\,$\lambda989$ line [contaminated in the right 
wing by the H$_2$ L11-0P(2) line from the \zabs = 3.025 DLA system --
see Fig.~5 in Levshakov et al. 2002b]. 
We do not detect high amplitude fluctuations in the density and
velocity fields in this compact system (see Fig.~10),
and as a result the observed metal line profiles are almost symmetric.
The low density region with the space coordinate $0 \leq x \leq 0.2$
does not contribute much to the line profiles, although a weak absorption
arising in this gas  can be seen in Fig.~9 in the \ion{C}{3}\,$\lambda$977 and 
\ion{C}{4}\,$\lambda1548$ lines at $v \simeq -30$ \kms. 

Since it is hardly possible that a cloud with a 120 pc size 
could exist in space on its own,
the two systems are probably physically related. 
For instance, this small and metal enriched cloud may be 
a condensation of a supernova-heated gas in a galactic halo 
seen in the \zabs = 2.96171 absorption lines.
This process known as a galactic fountain (Bregman 1980) is believed to give origin 
to the high metallicity HVCs observed in the halo of the Milky Way. 
The velocity excess of the HVCs is usually greater than 90 \kms\, 
which is consistent with 
the shift of $\simeq 120$ \kms\, 
between the redward absorption in the Ly$\alpha$ profile at 
\zabs = 2.96171 and the centre of the Ly$\alpha$ line at \zabs = 2.96591.

\subsubsection{Q0347--3819, \zabs = 2.98}

Although this system 
belongs to the most commonly observed type in the Ly$\alpha$ forest 
(apart from hydrogen lines only weak \ion{C}{4} and \ion{Si}{4} 
doublets and no apparent absorption 
in other ionic species are registered),
its hydrogen lines are unusually broad -- extended over 600 \kms.
  
Available  Ly$\alpha$, Ly$\beta$ and Ly$\gamma$ lines (see Fig.~11) allow us 
to estimate the hydrogen column density with a sufficiently high accuracy. 
According to the recovered physical parameters (see Table~2), in this case
we are dealing with a large ($L \simeq 50$ kpc) cloud of a 
rarefied ($n_0 \simeq 10^{-4}$ \cmm), metal-poor ($Z < 0.01\,Z_\odot$) and hot 
($T_{\rm kin} \simeq 40000$ K) gas.   
Blending of the \ion {C}{3}\,$\lambda977$ and \ion{Si}{3}\,$\lambda1206$ lines and 
the weakness of the \ion{Si}{4}\,$\lambda1393$ line 
do not allow us to estimate accurately
the mean ionization parameter. 
The $U_0$ value presented in Table ~2  should be considered as
a lower limit. If in reality $U_0$ is higher, then the absorber 
may have a lower mean
gas density and a larger linear size.

The most probable host for this system might be an external 
region of a giant galactic halo
or a large scale structure object.

\subsubsection{Q0347--3819, \zabs = 3.14}

The unsaturated Ly$\gamma$ line
gives accurate estimations of the total neutral hydrogen column density
(Fig.~12, Table~2). Clean continuum windows seen 
at the expected positions of metal lines make it possible to estimate the upper 
limits on metal
abundances and to calculate the total hydrogen column density $N_{\rm H}$.
The result obtained shows a rather low metallicity
cloud with [C/H] $< -2.2$ and the linear size $L > 13$ kpc. 
One may expect to observe similar systems in the outer parts of galactic halos.

\subsubsection{APM BR J0307--4945, \zabs = 4.81}

This is  the most distant absorber in our set
where a low metal abundance can be directly measured. 
Its hydrogen Ly$\alpha$ line is clearly seen at $\Delta v = 3000$ \kms\, 
in the wide emission blend Ly$\alpha$+\ion{N}{5}+\ion{Si}{2} shown in Fig.~1{\bf c}.
Absorption lines in Fig.~13 give the redshift \zabs = 4.8101 and, thus, 
this system has \zabs $> z_{\rm em}$. The same order of magnitude velocity
difference ($\Delta v \simeq 3000$ \kms)
between the H$_2$-bearing cloud at \zabs = 2.811 (Levshakov \& Varshalovich 1985)
and the quasar redshift $z_{\rm em} = 2.770$ (Foltz et al. 1988) 
has been observed toward
PKS 0528--250. This H$_2$ cloud seems to be at a distance larger than 10 kpc from the
quasar as shown by Srianand \& Petitjean (1998). 
However, in our case we are not able to
estimate the proximity of the quasar. We can only assume
that the photoionization of the \zabs = 4.8101 system
could be affected by the quasar radiation. 

This assumption is supported by the following facts. 
It turned out to be impossible to fit
all available lines with the HM ionizing spectrum: the relative intensities of 
the \ion{C}{3}\,$\lambda977$ and \ion{C}{4}\,$\lambda1548, 1550$ 
required very high $U_0$ values ($U_0 > 0.1$) which
contradicted with the shallow extended wings of the Ly$\alpha$ line.
On the contrary, the MF ionizing spectrum, corresponding to the AGN, 
allowed to fit all lines and delivered a self-consistent set of physical
parameters (Table~2, Figs.~13 and 14). The intensity at 1 Ry was set to
$J_{912} = 10^{-22}$ erg cm$^{-2}$ s$^{-1}$ Hz$^{-1}$ sr$^{-1}$ 
which corresponds to the
intensity of the HM spectrum at $z = 4.9$.

According to the recovered values, the absorber at \zabs = 4.81 
is a metal-poor cloud with the linear size of about 25 kpc and the mean
density $n_0 \simeq 2\times10^{-4}$ \cmm. Fig.~14 
shows that the shallow wings of the
Ly$\alpha$ line are produced by the streaming out
low density gas whereas the central region of the cloud remains very quiet.

Since we cleary see only the \ion{C}{4} doublet\footnote{The
$\lambda$1550 \AA\, component is slightly blended with the
night sky lines, whereas the $\lambda1548$ \AA\, line
is clear as found from the comparison of the two
spectra of the APM BR J0307--4945 taken in a two month interval.} 
and have upper limits for the intensity
of the \ion{C}{3}\,$\lambda977$ and \ion{Si}{4}\,$\lambda1393$ lines, the 
mean ionization parameter $U_0$ 
listed in Table~2 should be considered as a lower limit
implying that lower metal abundances and lower gas densities may also be possible.
Taking into account that the density $n_0$ and by this the
linear size $L$ scales with the intensity
of the radiation field (see eq.[1]), the value of $L$ becomes quite uncertain.
It could be larger or smaller than the estimated size of 25 kpc (this uncertainty is
marked by `?' in Table~2).

An absorption system with similar spectral characteristics 
(wide and shallow wings of the Ly$\alpha$ line, a weak \ion{C}{4}
doublet) was previously observed by Reimers et al. (2001) at \zabs = 1.674 toward
HE~0515--4414 ($z_{\rm em} = 1.73$). The system at \zabs = 1.674 shows
in addition a strong \ion{O}{6} doublet which, 
unfortunately, cannot be identified in
the \zabs = 4.81 cloud because of blending in the dense Ly$\alpha$ forest. 

Thus we may conclude that such systems are probably formed in the gas clouds
affected by the QSO radiation.

\subsection{Ly$\alpha$ systems with metallicity gradient}

\subsubsection{Q0347--3819, \zabs = 2.848 and 2.899}

These systems present several seemingly unblended hydrogen lines  
and pronounced lines of highly ionized silicon and carbon 
(Figs.~15, 16; Table~3). 
Additionally, the \ion{N}{5}\,$\lambda1238$ and \ion{O}{6}\,$\lambda1031$ lines
were identified at \zabs = 2.848. 

In spite of the multiple \ion{H}{1} lines, it turned out that
the MCI failed
to fit adequately all available ionic profiles in the apparent 
velocity ranges when homogeneous metallicities over the entire absorbing regions
were assumed.
The most sensitive restrictions in these calculations
are set by the continuum windows 
seen in the profiles of the strong \ion{C}{3}\,$\lambda977$ and 
\ion{Si}{3}\,$\lambda1206$ lines. 
We consider the observed lines from these systems as
arising in clouds with metallicity gradients for the following reasons.

The hydrogen Ly$\alpha$ and Ly$\beta$ lines seen at \zabs = 2.848 in the
$\Delta v$ range between 100 \kms\, and 400 \kms\, do not show any metal
absorption in this velocity interval. Absorption features seen in the range
140 \kms\,$\leq \Delta v \leq 400$ \kms\, in Fig.~15 (panel \ion{C}{3})
and in the range 180 \kms\,$\leq \Delta v \leq 400$ \kms\, in panel \ion{Si}{3}
cannot be attributed to the corresponding hydrogen sub-system, otherwise
we should observe pronounced \ion{C}{4} lines in the same velocity range.
If the redward hydrogen absorption is physically connected with the blueward one,
then we may conclude that the \zabs = 2.848 absorber has a metallicity gradient.

A very similar picture is seen in the \zabs = 2.899 system for the blueward portions
of the hydrogen lines in the range $-220$ \kms\,$\la \Delta v \la -80$ \kms
(Fig.~16).
The \ion{C}{3} $\lambda977$ line (which is the most sensitive absorption line
in a wide range of the ionization parameter from $\log U_0 \ga -4$ to 
$\log U_0 \simeq 0$) shows no absorption in this velocity range.

In principle, the \zabs = 2.899 system may be caused by a blending effect
when the line of sight intersects two separate clouds 
(one of them supposingly metal-free)
which have approximately the same radial velocities. As for the \zabs = 2.848 system,
a fortuitous blending  can be ruled out because we observe two absorption features
of approximately equal depth at $v \simeq -80$ \kms\, 
in the Ly$\alpha$ and \ion{C}{4}
profiles -- the configuration noticed previously 
at \zabs = 1.385 toward HE~0515--4414 by 
Reimers et al. (2001), i.e. this unusual pattern is not unique and seems to indicate
some special kind of absorption systems.
Preliminarily we may conclude that  the system at \zabs = 2.848 
as well as probably that at \zabs = 2.899 arise when the line of 
sight intersects a distant halo with very low (if any)
metal content and encounters a metal-rich HVC. 
Unfortunately, 
the accurate quantitative analysis of both systems cannot be carried out
because the 
velocity excesses of the assumed HVCs are not large enough to separate the inputs
to the hydrogen lines from the halos and from the clouds  
like it was possible in the case of the \zabs = 2.962 and \zabs = 2.965 systems. 
Therefore the results presented in Table~3
are tentative, but they are physically reasonable and self-consistent. 

The corresponding synthetic spectra are shown in Figs.~15 and 16 by the solid lines.  
It should be noted that for both systems, only
one wing of the available \ion{H}{1} lines (marked with the horizontal
bold lines) was included in the analysis, 
whereas the synthetic profiles for the entire 
\ion{H}{1} lines of the encountered HVCs were
computed using the velocity and density distributions estimated from the metal
profiles.

According to the data from Table~3, the suggested HVCs belong to different types.
The system at \zabs = 2.899 
has the metal abundance pattern, the set of the ions observed (notice
the pronouced \ion{Si}{4} doublet) and the size similar to those estimated for the
supposed 
HVC at \zabs = 2.965 and consistent with the parameters of the HVCs observed in the
Milky Way (see Appendix). The system at \zabs = 2.848 is more higher ionized
(\ion{O}{6} and \ion{N}{5} and no \ion{Si}{4}) and it has the size of
several kiloparsecs. Highly ionized HVCs with similar parameters 
were observed by Sembach et al. (1999) near the Milky Way. 
Their origin is still uncertain. Probably the \zabs = 2.848
absorber may belong to the intercluster gas clouds in a distant group of galaxies 
as it was suggested by Sembach et al. for the local highly ionized HVCs.

\section{Discussion}

\subsection{The origin of metal systems}

Metal systems with $N$(\ion{H}{1}) $< 5\times10^{16}$ \cm\,
are usually believed to   
originate in galactic halos at different galactocentric distances.
At low redshifts ($z < 1$) the galaxies associated with certain
metallic absorptions (e.g. \ion{C}{4}) can be in
most cases identified directly 
(e.g. Chen et al. 2001a). Our results on absorption systems with
$z \ga 2$ also support this assumption:
absorbers with \zabs = 1.87 (Paper~I),
2.54, 2.65, 2.962, 2.98 (present paper)
are produced by metal-enriched ($Z < 0.1\,Z_\odot$), hot ($T_{\rm kin} \ga 20000$ K),
rarefied ($n_0 \simeq 10^{-4} - 10^{-3}$ \cmm) gas clouds
which have typical linear sizes of $L > 10$~kpc.
These parameters are consistent with contemporary models of galactic halos
(e.g. Viegas, Friaca \& Gruenwald 1999).

The nature of Lyman limit absorbers is less understood.
Mo \& Miralda-Escud\'e (1996) associate them with 
cold photoionized clouds randomly
moving in hot spherical halos. The clouds are supposed to form from the
initial density inhomogeneities in the accreting intergalactic gas during its
cooling. Both the cloud and the halo  obviously reveal
the equal metallicity since they are formed from the same gas. 
In our study two LLSs with \zabs = 1.92 (Paper~I) and 2.81 (present paper)
can be related to the absorbers of this type.
 
However, this scenario obviously fails   
to explain metal abundant ($Z > 0.1\,Z_\odot$) systems since 
it is hard to understand 
how the whole halo can be metal-enriched to such a high level. 
It was shown by
hydrodynamic simulations (e.g. Katz et al. 1996; Gardner et al. 2001) 
that LLSs can also arise on lines of sight that pass through
small protogalaxies. We found two systems with \zabs = 1.94 (Paper~I)
and 4.21 (present paper) that can be explained within this framework. 
These metal-rich ($Z \simeq 1/3\,Z_\odot$) absorbers with the sizes of
several kpc are probably hosted by objects that may be akin to
the local compact blue galaxies.

Some absorbers in our present study (\zabs = 2.21, 2.965, and, possibly, 2.89)
reveal small linear sizes ($L < 1$ kpc) together with very high metal content
($Z \simeq 1/2\,Z_\odot$). These three systems may be explained in the framework
of the process known as a
galactic fountain~: metal-enriched (supernova-heated) gas arises from the inner
region of a galaxy and condenses into the clouds within the hot galactic halo.
After formation, clouds fall back toward the galaxy centre because of their
higher density. It is supposed 
that high-metallicity HVCs observed in the Milky Way halo
are formed by this mechanism (Bregman 1980).
The HVCs are common objects in our Galaxy and are detected in every longitude
and latitude region. If galactic 
fountain works also in distant galaxies, it would be 
quite probable to encounter such a cloud on the line of sight which intersects
the galactic halo, as also discussed by Charlton et al. (2001).
Another type of HVCs -- hot, highly ionized clouds with sizes of several kiloparsecs
 -- is represented by the absorption system
at \zabs = 2.848. The origin of this type of HVCs is uncertain, but they may be 
produced by the intergalactic metal-enriched gas falling onto metal-poor 
galactic halos.

Measured abundances of C and Si are depicted 
versus logarithmic sizes of the studied systems in Fig.~17.
Systematically higher metal abundances are seen in compact systems
with linear sizes $L < 4$ kpc.
This result seems to indicate that the more effective metal enrichment occurs
within relatively compact regions.

Our results show
that Lyman limit systems are a {\it heterogeneous} population
which is formed in at least three different environments. This should be
taken into account when statistics of LLSs is used to verify different
models in hydrodynamic cosmological simulations.

\subsection{$\sigma_{\rm v} - N_{\rm H}\,L$ relation}

If QSO metal systems are formed in gas clouds gravitationally bound with
intervening galaxies, the internal kinematics of the QSO absorbers should be
closely related to the total masses of the host galaxies. In case of galactic
population, different types of galaxies show different scaling relations
between the linear size 
and the velocity width of emission lines (e.g., Mall\'en-Ornelas
et al. 1999). Possible correlation between  
the absorber linear size $L$ and its 
line-of-sight velocity dispersion $\sigma_{\rm v}$
was also mentioned in Paper~I.

The correlation between $\sigma_{\rm v}$ and $L$ stems from the
virial theorem which states~:
\begin{equation}
\sigma^2_{\rm v} \sim \frac{M}{L} \sim n_0\,L^2 = N_{\rm H}\,L\; .
\label{eq:E1}
\end{equation}
Assuming that the gas systems are in quasi-equilibrium, 
one can expect $\sigma_{\rm v} \sim \sqrt{N_{\rm H}\,L}$.

In Fig.~18 we examine our systems by comparing their kinematics ($\sigma_{\rm v}$)
with measured sizes ($L$) and total gas column densities ($N_{\rm H}$). 
Shown are the data
for all QSO absorbers studied in Paper~I and in the present paper except for the
systems at \zabs = 2.848 and 2.899 (Table~3) which show inhomogeneous metallicities.
It is seen that 
in the $\log (\sigma_{\rm v})$ versus $\log (N_{\rm H}\,L)$ diagram, most
systems with linear sizes $L > 1$ kpc lie along the line with the slope 
$\kappa = 0.30\pm0.03$ (1 $\sigma$ c.l.). 

Taking into account that we know neither the impact parameters nor the halo
density distributions, this result can be considered as a quite good fit to
the relation (1). Hence we may conclude that most absorbers with $L > 1$ kpc are
gravitationally bound with systems that appear to be in virial
equilibrium at the cosmic time when the corresponding Ly$\alpha$ absorbers
were formed. The possible consequence of this conclusion is that since the most metal
rich absorbers identified in the QSO spectra arise in the galactic 
systems the question
whether the intergalactic matter is metal enriched or pristine remains still open.

\subsection{[$\alpha$-element/iron-peak] ratio}

The metal abundances measured in the \zabs = 2.21 LLS (Table~1)
can be used to estimate the $\alpha$-element 
to the iron-peak group ratio which is 
a good indicator of the chemical evolutionary status of high redshift gas clouds.
During the chemical evolution, heavy elements produced in stars
show different nucleosynthetic
histories so that their relative abundances vary with cosmic time.

Oxygen and other $\alpha$-chain elements are mainly produced by Type~II SNe,
while iron is also a product of Type~Ia SNe which have longer evolution scales.
In the early stages of the chemical evolution of galaxies ($\Delta t \la
2\times10^7$ yr) the interstellar gas is likely enriched by Type~II SNe products,
while at $\Delta t \ga 10^8$ yr, the [$\alpha$/Fe] ratio should decline.
Observations reveal both low
[e.g. $\simeq 0.1-0.2$ in the \zabs = 3.390 dust-free DLA
(Q0000--2621; Molaro et al. 2001) and in the
\zabs = 3.386 DLA (Q0201+1120; Ellison et al. 2000)], 
and high
[e.g. $\simeq 0.7$ in the DLA I~Zw 18 (Levshakov, Kegel \& 
Agafonova 2001) and $0.68\pm0.08$ in the \zabs = 3.025 DLA
(Q0347--3819; Levshakov et al. 2002b)]  
ratios of [$\alpha$-element/iron-peak]. 

Oxygen with its weak affinity with dust grains is a good tracer of the
$\alpha$-element abundances. Nevertheless, the intrinsic [$\alpha$/Fe] ratio may be
affected by depletion of iron since being a refractory element  
iron may be partly locked into dust grains. 
The dust content in the \zabs = 2.21 LLS may not, however, be too high.
The relative abundances of the $\alpha$-elements O, Mg and Si 
are [Si/O] $= -0.14\pm0.11$ and [Mg/O] = $-0.27\pm0.11$.
In Galactic stars the $\alpha$-elements show
the same behaviour relative to iron-peak elements 
(oversolar at [Fe/H] $\la -1$; see, e.g., Gaswami \& Prantzos 2000). 
We thus expect to find solar $\alpha$-element ratios in dust-free absorbing regions,
as observed, e.g.,
in the mentioned above \zabs = 3.390 DLA toward Q0000--2620.
A negative value of [Mg/O] found in this LLS may
indicate the presence of some amount of dust with a depletion factor of about
0.2 dex for the magnesium abundance.
If, however, only the gas-phase abundances of O and Fe are taken, 
the upper bound on the [O/Fe] ratio is $0.65\pm0.11$, which is comparable with
that found, for instance, 
in the \zabs = 3.025 DLA toward Q0347--3819 where the dust-to-gas ratio is
$\simeq 1/30$ of the mean 
Galactic interstellar medium value (Levshakov et al. 2002b).
The enrichment of the $\alpha$-elements in the \zabs = 2.21 LLS is
also supported by the relative abundances of Si, Mg to Fe:
[Si/Fe] = $0.51\pm0.11$ and [Mg/Fe] = $0.38\pm0.11$. Thus, the absorbing cloud
at \zabs = 2.21 appears to be a chemically young object. 

\section{Summary}

We have deduced the physical properties of ten
absorption-line systems in the range $\Delta z = 2.21 - 2.966$ toward
Q0347--3819 and of two systems at \zabs = 4.21 and 4.81 toward APM BR J0307--4945. 
The main conclusions are as follows~:
\begin{enumerate}
\item The analyzed
Lyman limit systems belong to a {\it heterogeneous} population
which is formed by at least three groups of absorbers~: 
($i$) extended metal-poor gas 
halos of distant galaxies; 
($ii$) gas in dwarf galaxies;
($iii$) metal-enriched gas arising from the inner
galactic regions and condensing into the clouds within the hot galactic halo
(galactic fountain).
While the interpretation of a single system is sometimes subject to large 
uncertainties as discussed in chapter 4, the existence of a wide spread of
properties in the different systems is certainly proved.
\item A correlation between the line-of-sight velocity dispersion $\sigma_{\rm v}$
and the linear size $L$ of the absorbing systems noted in Paper~I is confirmed.  
New results show that 
large size QSO absorbers ($L > 1$ kpc)
obey a scaling law $\sigma_{\rm v} \sim (N_{\rm H}\,L)^{0.3}$ over
two decades in velocity dispersion and 
in the product $(N_{\rm H}\,L)$. This means that
the majority of the 
metal absorbers are probably bound to the galactic systems and, hence,
the question whether the IGM is enriched or pristine requires further investigations.
\item Systematically higher metal abundances are found in compact systems:
in our sample there are no small size systems ($L < 1$ kpc) with metallicity
lower than $0.1\,Z_\odot$.
\item The gas-phase metal abundances from the \zabs = 2.21 LLS reveal a pronounced
[$\alpha$-element/iron-peak] enhancement with [O/Fe]  = $0.65\pm0.11$ at the
$6\sigma$ confidence level, the first time when this abundance pattern
is unambiguously found in a LLS. The measured [O/Fe] ratio implies that the
chemical history of this LLS is $\la 10^8$ yr.
\item Absorption system at \zabs = 2.21 and 2.965 
and possible the systems at  \zabs = 2.848 
and 2.899 toward Q0347--3819 show characteristics very similar to that observed
for different types of HVCs in the Milky Way 
and may be interpreted as being the high-redshift counterparts
of these Galactic objects. 
\end{enumerate}

\acknowledgments

We thank Prof. W. H. Kegel for helpful correspondence and our anonymous referee for
many helpful suggestions.
S.A.L. gratefully acknowledges the hospitality of 
the National Astronomical Observatory
of Japan (Mitaka) where this work was performed. 
The work of S.A.L. and I.I.A. is supported in part by the 
RFBR grant No.~00-02-16007.

\appendix

\section{Possible counterparts of QSO absorbers}

In our interpretation of the nature of the different metal systems with
$N$(\ion{H}{1}) $\simeq 10^{14}-10^{17}$ \cm\, 
we have referred to a set of absorbers whose physical parameters are summarized
in Table~4. 
Comparison between the reference absorbers and the QSO systems is based on
following parameters~:
\begin{enumerate}
\item Linear sizes. The systems can be divided into small-size
($L < 1$ kpc), intermediate-size (1 kpc $\la L \la 10$ kpc), and
large-size (10 kpc $\la L \la 150$ kpc) absorbing regions connected
with galactic gaseous envelopes, as well as into very large-size
($L \ga$ 150 kpc) absorbers (filaments) showing correlation with the
large-scale distribution of galaxies (e.g. Penton et al. 2002).
Filaments are probably intergalactic material not recycled by galaxies.
\item Metallicities. Values of $0.3 \la Z/Z_\odot \la 1$,
$0.03 \la Z/Z_\odot \la 0.3$, and $Z/Z_\odot < 0.03$
classify the systems as metal rich, metal enriched, and metal poor,
respectively.
\item Metallicity patterns. Relative element abundances
allow us to distinguish between
chemically young and old systems (low and high 
relative [$\alpha$-element/iron-peak] ratio, respectively).
\end{enumerate}

\clearpage
\begin{figure}
\figcaption[f1.ps]{
Emission line profiles in the VLT/UVES spectrum of APM BR J0307--4945. 
The zero velocity
corresponds to the redshift $z_{\rm em} = 4.7525$ 
and indicates the adopted line centre of
the \ion{O}{1} $\lambda1304$ line. 
A smooth fit to the \ion{C}{4} $\lambda1549$ profile
superimposed on the original line is shown in panel ({\bf b}). 
Comparison of the synthetic
\ion{C}{4} profile scaled to match other emission lines is illustrated in panels
({\bf a}), ({\bf c}), ({\bf d}), and ({\bf e}). 
A combined profile constructed from the same
synthetic \ion{C}{4} profile properly scaled 
for each emission line and shifted for the
\ion{N}{5} $\lambda1240$ and \ion{Si}{2} $\lambda1264$ lines to their centers 
(assuming $z_{\rm em} = 4.7525$) is superposed
on the whole extended emission line in panel ({\bf f}).
\label{fig1}}
\end{figure}

\begin{figure}
\figcaption[f2.ps]{
Hydrogen and metal absorption lines associated with the
\zabs = 2.21 LLS toward Q0347--3819
(normalized intensities are shown by dots with $1\sigma$ error bars).
The zero radial velocity is fixed at $z = 2.21398$. Smooth lines are
the synthetic spectra convolved with the corresponding point-spread
functions and computed with the physical
parameters from Table~1. Bold horizontal lines mark pixels included
in the optimization procedure. Spectra obtained with the VLT/UVES
and the Keck/HIRES spectrographs are marked by VLT and Keck, respectively.
The normalized $\chi^2_{\rm min} = 1.39$ (the number of degrees of
freedom $\nu = 1255$).
\label{fig2}}
\end{figure}

\begin{figure}
\figcaption[f3.ps]{
Computed velocity (upper panel) and gas density (lower panel) distributions
along the line of sight within the \zabs = 2.21 LLS toward Q0347--3819.
Shown are patterns rearranged according to the principle of minimum entropy
production rate (see Paper~I).
\label{fig3}}
\end{figure}

\begin{figure}
\figcaption[f4.ps]{
Same as Fig.~2 but for the \zabs = 2.81 LLS.
The zero radial velocity is fixed at $z = 2.8102$.
The corresponding physical parameters are listed in Table~1.
$\chi^2_{\rm min} = 1.38$,  $\nu = 763$.
\label{fig4}}
\end{figure}

\begin{figure}
\figcaption[f5.ps]{
Same as Fig.~2 but for the \zabs = 4.21 LLS toward APM BR J0307--4945.
The zero radial velocity is fixed at $z = 4.2144$. 
The physical parameters listed in Table~1, column 4,
correspond to the system centered at $v \simeq 200$ \kms. The synthetic
spectra for this system are shown by the solid lines 
($\chi^2_{\rm min} = 1.83$, $\nu = 1436$). 
The synthetic spectra shown by the dotted lines are calculated with the parameters
listed in Table~1, column 5, which are
obtained from the analysis of the entire system from $v = -300$ \kms\, to 400 \kms\,
($\chi^2_{\rm min} = 2.61$, $\nu = 2325$). 
\label{fig5}}
\end{figure}

\begin{figure}
\figcaption[f6.ps]{
Same as Fig.~2 but for the \zabs = 2.53 system.
The zero radial velocity is fixed at $z = 2.5370$. 
The corresponding physical parameters are listed in Table~2.
$\chi^2_{\rm min} = 1.07$, $\nu = 602$.
\label{fig6}}
\end{figure}

\begin{figure}
\figcaption[f7.ps]{
Same as Fig.~2 but for the \zabs = 2.65 system.
The zero radial velocity is fixed at $z = 2.65044$. 
The corresponding physical parameters are listed in Table~2.
$\chi^2_{\rm min} = 1.15$, $\nu = 962$.
\label{fig7}}
\end{figure}

\begin{figure}
\figcaption[f8.ps]{
Same as Fig.~2 but for the \zabs = 2.962 system.
The zero radial velocity is fixed at $z = 2.96171$. 
The corresponding physical parameters are listed in Table~2.
$\chi^2_{\rm min} = 1.20$, $\nu = 627$.
\label{fig8}}
\end{figure}

\begin{figure}
\figcaption[f9.ps]{
Same as Fig.~2 but for the \zabs = 2.966 system.
The zero radial velocity is fixed at $z = 2.96591$. 
The corresponding physical parameters are listed in Table~2.
$\chi^2_{\rm min} = 1.10$, $\nu = 149$.
\label{fig9}}
\end{figure}

\begin{figure}
\figcaption[f10.ps]{
Computed velocity (upper panel) and gas density (lower panel) distributions
along the line of sight within the \zabs = 2.966 absorber toward Q0347--3819.
Shown are patterns rearranged according to the principle of minimum entropy
production rate (see Paper~I).
\label{fig10}}
\end{figure}

\begin{figure}
\figcaption[f11.ps]{
Same as Fig.~2 but for the \zabs = 2.98 system.
The zero radial velocity is fixed at $z = 2.97915$. 
The corresponding physical parameters are listed in Table~2.
$\chi^2_{\rm min} = 1.20$, $\nu = 858$.
\label{fig11}}
\end{figure}

\begin{figure}
\figcaption[f12.ps]{
Same as Fig.~2 but for the \zabs = 3.14 system.
The zero radial velocity is fixed at $z = 3.13985$. 
The corresponding physical parameters are listed in Table~2.
$\chi^2_{\rm min} = 1.00$, $\nu = 407$.
\label{fig12}}
\end{figure}

\begin{figure}
\figcaption[f13.ps]{
Same as Fig.~2 but for the \zabs = 4.81 Ly$\alpha$ system toward APM BR J0307--4945.
The zero radial velocity is fixed at $z = 4.8101$. 
The corresponding physical parameters are listed in Table~2.
$\chi^2_{\rm min} = 0.85$, $\nu = 303$.
\label{fig13}}
\end{figure}

\begin{figure}
\figcaption[f14.ps]{
Computed velocity (upper panel) and gas density (lower panel) distributions
along the line of sight within the \zabs = 4.81 absorber toward APM BR J0307--4945.
Shown are patterns rearranged according to the principle of minimum entropy
production rate (see Paper~I).
\label{fig14}}
\end{figure}

\begin{figure}
\figcaption[f15.ps]{
Same as Fig.~2 but for the \zabs = 2.84 Ly$\alpha$ 
system with inhomogeneous metallicity.
The zero radial velocity is fixed at $z = 2.84829$. 
The corresponding physical parameters are listed in Table~3.
$\chi^2_{\rm min} = 1.29$, $\nu = 284$.
\label{fig15}}
\end{figure}

\begin{figure}
\figcaption[f16.ps]{
Same as Fig.~2 but for the \zabs = 2.90 Ly$\alpha$ 
system with inhomogeneous metallicity.
The zero radial velocity is fixed at $z = 2.89922$. 
The corresponding physical parameters are listed in Table~3.
$\chi^2_{\rm min} = 1.35$, $\nu = 820$.
\label{fig16}}
\end{figure}

\begin{figure}
\figcaption[f17.ps]{
Abundances of carbon (circles) and silicon (squares)
plotted against the logarithmic linear 
size of the corresponding systems as measured by the
MCI procedure. 
Shown are the data from Table~1 (Paper~I) and form Tables~1 and 2 (present
study).  The data for the \zabs = 4.21 
system toward APM BR J0307--4945 are taken from column 4,
Table~1. 
The point at log (L) = 3.40 corresponds to the \zabs = 4.81 system toward
APM BR J0307--4945 where the linear size is very uncertain.
Small size QSO absorbers seem to have systematically higher metal content
than systems with larger sizes.
\label{fig17}}
\end{figure}

\begin{figure}
\figcaption[f18.ps]{
Plot of the line of sight velocity dispersion $\log (\sigma_{\rm v})$ vs. 
$\log (N_{\rm H}\,L)$ for
the sample of metal absorption-line systems. 
The dashed line is the linear regression $\log (\sigma_{\rm v}) \propto 
\kappa\,\log (N_{\rm H}\,L)$ calculated for the points shown as filled circles
having both horizontal error bars.
The slope $\kappa$ is equal to $0.30\pm0.03$ (1 $\sigma$ c.l.).
Shown are the data from Table~1 (Paper~I) and form Tables~1 and 2 (present
study). The data for the \zabs = 4.21 
system toward APM BR J0307--4945 are taken from column 4,
Table~1. 
The point at log ($N_{\rm H} L) = 41.08$ 
corresponds to the \zabs = 4.81 system toward
APM BR J0307--4945 where the linear size is very uncertain.
Open squares represent HVC-like systems
(see text for details).
\label{fig18}}
\end{figure}

\clearpage
\begin{deluxetable}{lcccc}
\tabletypesize{\scriptsize}
\tablecaption{
{\sc Physical parameters of the Lyman limit systems toward
Q0347--3819 and APM BR J0307--4945 derived by the MCI procedure
} 
\label{tbl-1}}
\tablewidth{0pt}
\tablehead{
\colhead{ } & \colhead{Q0347--3819} & \colhead{Q0347--3819} & 
\multicolumn{2}{c}{APM BR J0307--4945}\\ 
\colhead{Parameter$^a$} & \colhead{\zabs = 2.21398} & \colhead{\zabs = 2.8102} & 
\multicolumn{2}{c}{\zabs = 4.2144}\\ 
\colhead{(1)} & \colhead{(2)} & \colhead{(3)} & \colhead{(4)} & \colhead{(5)}
}
\startdata
$U_0$ & 2.1E-3 & 4.0E-2 & 7.5E-3 & 8.1E-3\\
$N_{\rm H}$, \cm & 2.2E19 & 8.3E19 & 4.2E19 & 9.1E19  \\
$\sigma_{\rm v}$, \kms & 80.8 & 59.5 & 80.0 & 170.0\\
$\sigma_{\rm y}$ & 1.0 & 1.15 & 1.2 & 1.25 \\
$Z^b_{\rm C}$ & 2.0E-4 & 3.1E-5 & 1.3E-4 & 2.0E-4\\ 
$Z_{\rm N}$ & $\ldots$ & $<$4.0E-6 & $<$2.5E-5 & $<$2.5E-5\\
$Z_{\rm O}$ & 4.4E-4 & $\ldots$ & $\ldots$ & $\ldots$ \\
$Z_{\rm Mg}$ & 1.5E-5 & $\ldots$ & $\ldots$ & $\ldots$ \\
$Z_{\rm Al}$ & $\ldots$ & $\ldots$ & $<$2.5E-6 & $<$2.5E-6\\
$Z_{\rm Si}$ & 2.0E-5 & 4.2E-6 & 1.1E-5 & 1.5E-5\\
$Z_{\rm Fe}$ & 5.0E-6 & $\ldots$ & $\ldots$ & $\ldots$\\
$[Z_{\rm C}]^c$ & $-0.28$ & $-1.10$ & $-0.47$ & $-0.27$\\
$[Z_{\rm N}]$ & $\ldots$ & $<-1.3$ & $<-0.5$ & $<-0.5$\\
$[Z_{\rm O}]$ & $-0.10$ & $\ldots$ & $\ldots$ & $\ldots$\\
$[Z_{\rm Mg}]$ & $-0.37$ & $\ldots$ & $\ldots$ & $\ldots$\\
$[Z_{\rm Al}]$ & $\ldots$ & $\ldots$ & $<-0.5$ & $<-0.5$\\
$[Z_{\rm Si}]$ & $-0.24$ & $-0.90$ & $-0.51$ & $-0.36$\\
$[Z_{\rm Fe}]$ & $-0.75$ & $\ldots$ & $\ldots$ & $\ldots$\\
$N$(H\,{\sc i}), \cm & 4.6E17 & 5.5E16 & 1.4E17 & 2.3E17\\ 
$N$(O\,{\sc i}) & 1.5E14 & $\ldots$ & $\ldots$ & $\ldots$\\
$N$(C\,{\sc ii}) & 1.5E15 & 5.2E13 & 4.8E14 & 1.2E15\\
$N$(Mg\,{\sc ii})& 1.1E14 & $\ldots$ & $\ldots$ & $\ldots$\\
$N$(Si\,{\sc ii})& 2.3E14 & 8.2E12 & 4.8E13 & 1.0E14\\
$N$(Fe\,{\sc ii})& 1.3E13 & $\ldots$ & $\ldots$ & $\ldots$\\
$N$(C\,{\sc iii})& $\ldots$ & 1.6E15 & 4.3E15 & 1.3E16\\
$N$(N\,{\sc iii})& $\ldots$ & $<$2.0E14 & $<$6.3E14 & $<$1.6E15\\
$N$(Al\,{\sc iii})& $<$8.1E12 & $\ldots$ & $<$1.1E13 & $<$3.4E13\\
$N$(Si\,{\sc iii})& 2.0E14 & 1.2e14 & 2.6E14 & 6.8E14\\
$N$(C\,{\sc iv})& 7.7E13 & $\ldots$ & 5.6E14 & 2.0E15\\
$N$(Si\,{\sc iv})& 3.0E13 & 6.7E13 & 9.0E13 & 2.9E14\\
$n_0$, \cmm & 2.0E-2 & 9.2E-4 & 2.5E-3 & 2.5E-3\\
$\langle T \rangle$, K & 9.1E3 & 2.4E4 & 1.2E4 & 1.2E4\\
$T_{\rm min}$ & 8.6E3 & 1.7E4 & 1.1E4 & 1.1E4\\
$T_{\rm max}$ & 9.8E3 & 3.3E4 & 1.4E4 & 1.5E4\\
$L$, kpc & 0.38 & 30 & 6.0 & 13\\
\enddata
\tablecomments{$^a$The internal errors of $U_0$, $N_{\rm H}$, $\sigma_{\rm  v}$, 
$\sigma_{\rm y}$ and $Z_{\rm a}$  are $\simeq 15-20$\%, the 
column density errors are $\la 10$\%, whereas 
$n_0$ and $L$ are known with
$\simeq 50$\% accuracy;\, $^bZ_{\rm X}$ = X/H;\,
$^c[Z_{\rm X}] = \log (Z_{\rm X}) - \log (Z_{\rm X})_\odot$\,
[solar abundances are taken from Holweger (2001) except Al for which
solar abundance from Grevesse \& Sauval (1998) is used ];
}
\end{deluxetable}

\clearpage
\begin{deluxetable}{lccccccc}
\tabletypesize{\scriptsize}
\tablecaption{
{\sc Physical parameters of systems 
with $N$(H\,{\sc i}) $< 5\times10^{16}$ \cm\, toward
Q0347--3819 (A) and APM BR J0307--4945 (B) derived by the MCI procedure
} 
\label{tbl-2}}
\tablewidth{0pt}
\tablehead{
\colhead{ } & \colhead{A, \zabs =} & \colhead{A, \zabs =} 
& \colhead{A, \zabs =} & \colhead{A, \zabs =} & \colhead{A, \zabs =} & 
\colhead{A, \zabs =} 
& \colhead{B, \zabs =}\\ 
\colhead{Parameter$^a$} & 
\colhead{2.5370} & 
\colhead{2.65044} & 
\colhead{2.96171} &
\colhead{2.96591} &
\colhead{2.97915} &
\colhead{3.13985} &
\colhead{4.8101} \\
\colhead{(1)} & \colhead{(2)} & \colhead{(3)} & \colhead{(4)} &
\colhead{(5)} & \colhead{(6)} &  \colhead{(7)} & \colhead{(8)} 
}
\startdata
$U_0$ & 2.9E-2 & 4.2E-2 & 8.5E-2 & 2.0E-2 & 0.15 & 0.11 & 3.8E-2\\
$N_{\rm H}$, \cm & 4.4E19 & 1.9E19 & 2.0E19 & 5.1E17 & 3.6E19 & 1.2E19 & 1.6E19\\
$\sigma_{\rm v}$, \kms & 68.5 & 73.0 & 88.0 & 11.6 & 110.8 & 25.3 & 33.8\\
$\sigma_{\rm y}$ & 1.05 & 0.7 & 0.85 & 0.92 & 1.12 & 1.0 & 0.87\\
$Z^b_{\rm C}$ & 5.0E-6 & 3.7E-5 & 2.0E-6 & 1.5E-4 & 3.1E-6 & $<$ 1.8E-6 & 3.3E-6\\ 
$Z_{\rm N}$ & $\ldots$ & $\ldots$ & $\ldots$ & 1.8E-5 & $\ldots$ & $\ldots$ & $\ldots$ \\
$Z_{\rm Si}$ & 1.2E-6 & 1.2E-5 & 1.1E-6 & 3.2E-5 & $<$3.5E-6 & $<$ 9.0E-7 & $<7.0$E-7 \\
$[Z_{\rm C}]^c$ & $-1.89$ & $-1.02$ & $>-2.3$ & $-0.42$ & $-2.07$ & $<-2.3$ & $<-2.0$ \\
$[Z_{\rm N}]$ & $\ldots$ & $\ldots$ & $\ldots$ & $-0.67$ & $\ldots$ & $\ldots$ & $\ldots$ \\
$[Z_{\rm Si}]$ & $-1.45$ & $-0.47$ & $-1.49$ & $-0.03$ & $<-1.0$ & $<-1.6$ & $<-1.7$ \\
$N$(H\,{\sc i}), \cm & 2.2E16 & 6.2E15 & 2.0E15 & 3.6E14 & 1.9E15 & 8.7E14 & 3.7E15\\ 
$N$(C\,{\sc ii}) & $\ldots$ & $\ldots$ & $\ldots$ & 9.4E11 & $\ldots$ & $<$4.7E10 & $<$1.5E11 \\
$N$(Si\,{\sc ii})& $\leq$1.2E12 & $\ldots$ & $\ldots$ & $<$1.8E11 & $\ldots$ & $<$1.6E10 & $\ldots$ \\
$N$(C\,{\sc iii})& $\ldots$ & $\ldots$ & 1.1E13 & 4.3E13 & 1.8E13 & $<$4.6E12 & $<4.8$E12\\
$N$(N\,{\sc iii})& $\ldots$ & $\ldots$ & $\ldots$ & 5.2E12 & $\ldots$ & $\ldots$  &  $\ldots$ \\
$N$(Si\,{\sc iii})& 1.6E13 & 2.7E13 & 1.1E12 & 3.6E12 & $<$9.1E11 & $<$2.8E11 & $\ldots$\\
$N$(C\,{\sc iv})& 4.7E13 & 2.5E14 & 8.4E12 & 1.9E13 & 2.0E13 & $<$4.5E12 & 9.4E12\\
$N$(Si\,{\sc iv})& 7.3E12 & 3.7E13 & 8.5E11 & 5.2E12 & $<$1.4E12 & $<$3.0E11 & $<6.7$E11\\
$n_0$, \cmm & $<$1.1E-3 & 5.7E-4 & $<$3.2E-4 & 1.5E-3 & $<$2.4E-4 & $<$3.1E-4 & 2.3E-4?\\
$\langle T \rangle$, K & 2.8E4 & 1.8E4 & 4.0E4 & 1.0E4 & 4.4E4 & 3.4E4 & 3.2E4\\
$T_{\rm min}$ & 1.9E4 & 1.4E4 & 2.6E4 & 7.0E3 & 3.0E4 & 2.7E4 & 2.2E4\\
$T_{\rm max}$ & 4.3E4 & 2.3E4 & 5.0E4 & 2.1E4 & 7.0E4 & 5.3E4 & 4.3E4\\
$L$, kpc & $>$13 & 11 & $>$20 & 0.12 & $>$50 & $>$13 & 25?\\
\enddata
\tablecomments{$^a$The internal errors of $U_0$, $N_{\rm H}$, $\sigma_{\rm  v}$, 
$\sigma_{\rm y}$ and $Z_{\rm a}$  are $\simeq 15-20$\% , the 
column density errors are $\la 10$\%, whereas 
$n_0$ and $L$ are known with
$\simeq 50$\% accuracy;\, $^bZ_{\rm X}$ = X/H;\,
$^c[Z_{\rm X}] = \log (Z_{\rm X}) - \log (Z_{\rm X})_\odot$\,
(solar abundances are taken from Holweger 2001);
}
\end{deluxetable}

\clearpage
\begin{deluxetable}{lcc}
\tabletypesize{\scriptsize}
\tablecaption{
{\sc Physical parameters of the Q0347--3819 Ly$\alpha$ systems with metallicity
gradient derived by the MCI procedure
} 
\label{tbl-3}}
\tablewidth{0pt}
\tablehead{
\colhead{Parameter$^a$} & 
\colhead{\zabs = 2.84829} & 
\colhead{\zabs = 2.89922} \\ 
\colhead{(1)} & \colhead{(2)} & \colhead{(3)} 
}
\startdata
$U_0$ & 0.1 & 2.2E-2 \\
$N_{\rm H}$, \cm & 2.3E18 & 1.9E18 \\
$\sigma_{\rm v}$, \kms & 34.9 & 54.3 \\
$\sigma_{\rm y}$ & 0.9 & 1.3 \\
$Z^b_{\rm C}$ & 9.1E-5 & 9.6E-5  \\ 
$Z_{\rm N}$  & $\la$2.5E-5 & $\ldots$ \\
$Z_{\rm O}$  & 3.5E-4 & $\ldots$ \\
$Z_{\rm Si}$ & $<$2.5E-5 & 2.4E-5 \\
$[Z_{\rm C}]^c$ & $-0.67$ & $-0.60$ \\
$[Z_{\rm N}]$   & $\la -0.5$ & $\ldots$ \\
$[Z_{\rm O}]$   & $-0.21$ & $\ldots$ \\
$[Z_{\rm Si}]$ & $<-0.14$ & $-0.15$ \\
$N$(H\,{\sc i}), \cm & 1.8E14 & 4.1E15 \\ 
$N$(C\,{\sc ii}) & $\ldots$ & 1.1E13 \\
$N$(Si\,{\sc ii})& $\ldots$ & 3.6E12 \\
$N$(C\,{\sc iii})& 2.7E13 & 1.4E14  \\
$N$(Si\,{\sc iii})& $\ldots$ & 2.3E13 \\
$N$(C\,{\sc iv})& 5.8E13 & 3.0E13  \\
$N$(Si\,{\sc iv})& $<$1.3E12 & 1.3E13 \\
$N$(N\,{\sc v}) & $\la$ 8.3E12 & $\ldots$ \\
$N$(O\,{\sc vi}) & 5.3E13 & $\ldots$ \\
$n_0$, \cmm & 2.5E-4 & 2.0E-3 \\
$\langle T \rangle$, K & 2.3E4 & 1.2E4  \\
$T_{\rm min}$ & 1.9E4 & 9.9E3 \\
$T_{\rm max}$ & 2.9E4 & 1.4E4 \\
$L$, kpc & 3.0 & 0.33 \\
\enddata
\tablecomments{$^a$The internal errors of $U_0$, $N_{\rm H}$, $\sigma_{\rm  v}$, 
$\sigma_{\rm y}$ and $Z_{\rm a}$  are $\simeq 15-20$\% , the 
column density errors are $\la 10$\%, whereas 
$n_0$ and $L$ are known with
$\simeq 50$\% accuracy;\, $^bZ_{\rm X}$ = X/H;\,
$^c[Z_{\rm X}] = \log (Z_{\rm X}) - \log (Z_{\rm X})_\odot$\,
(solar abundances are taken from Holweger 2001);
}
\end{deluxetable}

\clearpage
\begin{deluxetable}{lcclc}
\tabletypesize{\scriptsize}
\tablecaption{
{\sc Parameters of absorbers which may give rise to 
QSO non-DLA systems
} 
\label{tbl-A1}}
\tablewidth{0pt}
\tablehead{
\colhead{Type of Absorber} & 
\colhead{Scale Size,} & 
\colhead{Metallicity,} &
\colhead{Metallicity} & 
\colhead{Ref.} \\ 
\colhead{ } & 
\colhead{$L$, kpc} & 
\colhead{$X = Z/Z_\odot$} &
\colhead{pattern$^a$} &
\colhead{ }  
}
\startdata
1. HVC-type absorbers inside&  &  
& [C/Fe]\,$\simeq0.2$, [N/Fe]\,\,\,\,$\simeq0.6$ & 1,2\\
\hspace*{0.4cm}galactic halos & $L < 1$ &
$0.1 \la X \la 1$ & [O/Fe]\,$\simeq0.7$, [Mg/Fe]$\simeq0.2$ &  \\  
 & & & [Al/Fe]$\simeq0.3$, [Si/Fe]\,\,\,$\simeq0.3$ &  \\
2. Gas in dwarf galaxies of & & & & \\
\hspace*{0.4cm}($i$)\,\, low-metallicity &$1 \la L \la 10$ &$0.03 \la X \la 0.06$ &
[C/Fe]$\simeq-0.2$, [O/Fe]$\simeq0.3$ & 3,4\\
 & &  &[N/Fe]$\simeq-0.2$, [Si/Fe]$\simeq0.1$   &  \\
\hspace*{0.4cm}($ii$) high-metallicity &$1 \la L \la 10$ &$0.06 < X \la 0.3$ &
[C/Fe]$\simeq\,\,\,\, 0.2$,\, [O/Fe]$\simeq0.4$ &3,4 \\
 & &  &[N/Fe]$\simeq-0.1$, [Si/Fe]$\simeq0.3$   &  \\
3. Gaseous envelopes$^b$  & & & &  \\
\hspace*{0.4cm}of galaxies seen in & & & & \\
\hspace*{0.4cm}($i$)\,\,\,\,\,\, H\,{\sc i}\, Ly$\alpha$&$\,\,\,10h^{-1} \la L \la 160h^{-1}$ 
&  & & 5,10 \\
\hspace*{0.4cm}($ii$)\,\,\,\, Mg\,{\sc ii}\,\,$\lambda\lambda 2796, 2803$ &
$15h^{-1} \la L \la 75h^{-1}$ &$X < 1$ & & 6,7\\
\hspace*{0.4cm}($iii$)\,\, C\,{\sc iv}\,\,$\lambda\lambda 1548, 1550$ &
$100h^{-1} \la L \la 180h^{-1}$ &$X < 1$ & & 8,9\\
\\
4. Large-scale filaments & & & &  \\
\hspace*{0.4cm}around galaxies & $L \ga 150h^{-1}$ &$X \ll 1$ & 
[Si/C]$\simeq0.4$,\, [N/C]$\simeq-0.7$ & 10
\enddata
\tablecomments{$^a$The patterns for the absorbers of type 1 and 2 
are taken from  Savage \& Sembach (1996) and
Izotov \& Thuan (1999), respectively, whereas the pattern for the case 4
is a characteristic chemical abundance of low-metallicity intergalactic
systems (Songaila 1998). 
$^b$Following Chen et al. (1998), by `gaseous envelopes' we mean a gaseous
structure of large covering factor but unspecified geometry or filling factor.\\
$^1$Savage \& Sembach 1996.\\
$^2$Wakker 2001.\\
$^3$Izotov \& Thuan 1999.\\
$^4$Mall\'en-Ornelas et al. 1999.\\
$^5$Chen et al. 1998.\\
$^6$Bergeron \& Boiss\'e 1991.\\
$^7$Steidel et al. 1997.\\
$^8$Chen et al. 2001a.\\
$^9$Chen et al. 2001b.\\
$^{10}$Penton, Stocke, \& Shull 2002. 
}
\end{deluxetable}

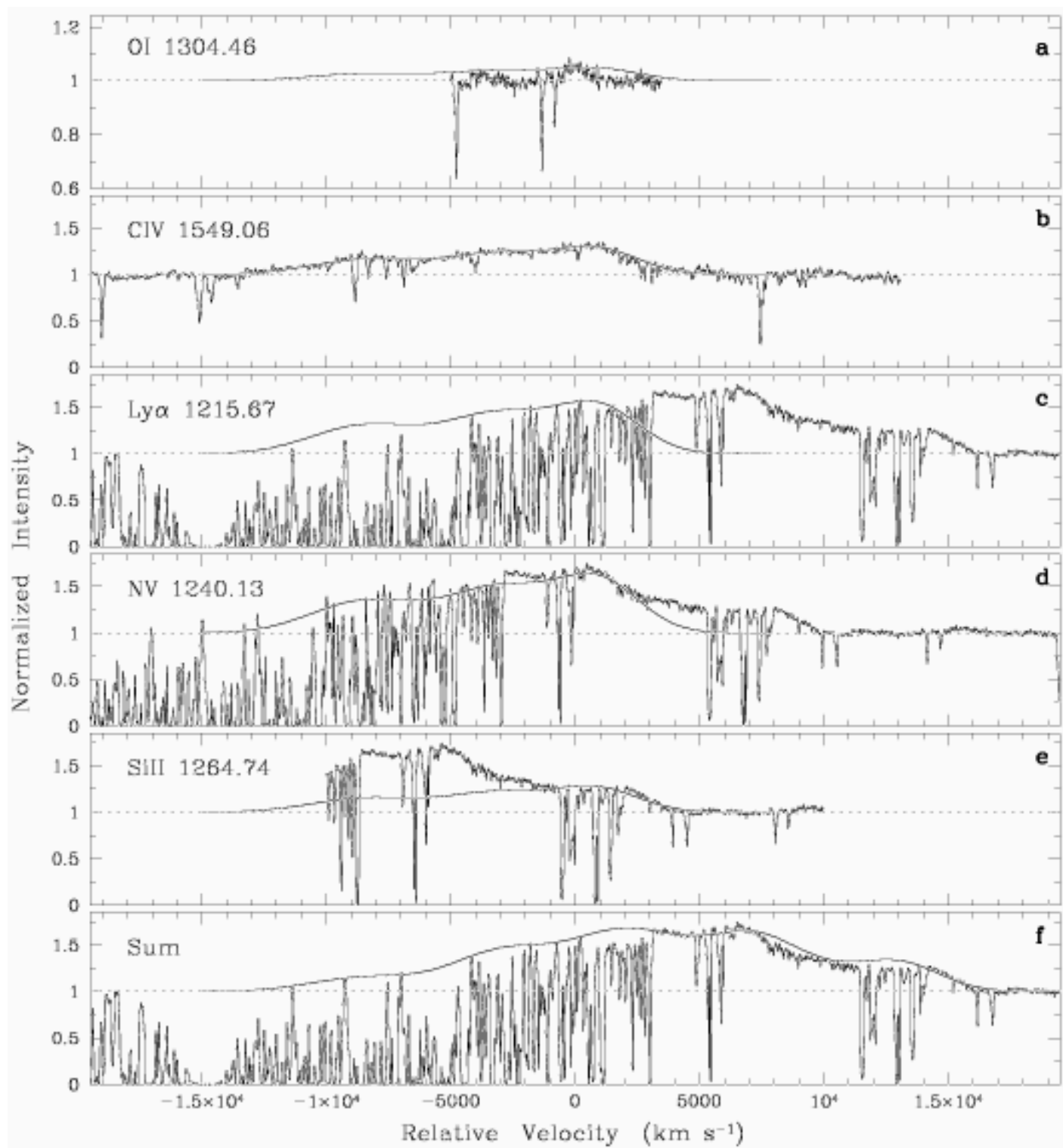
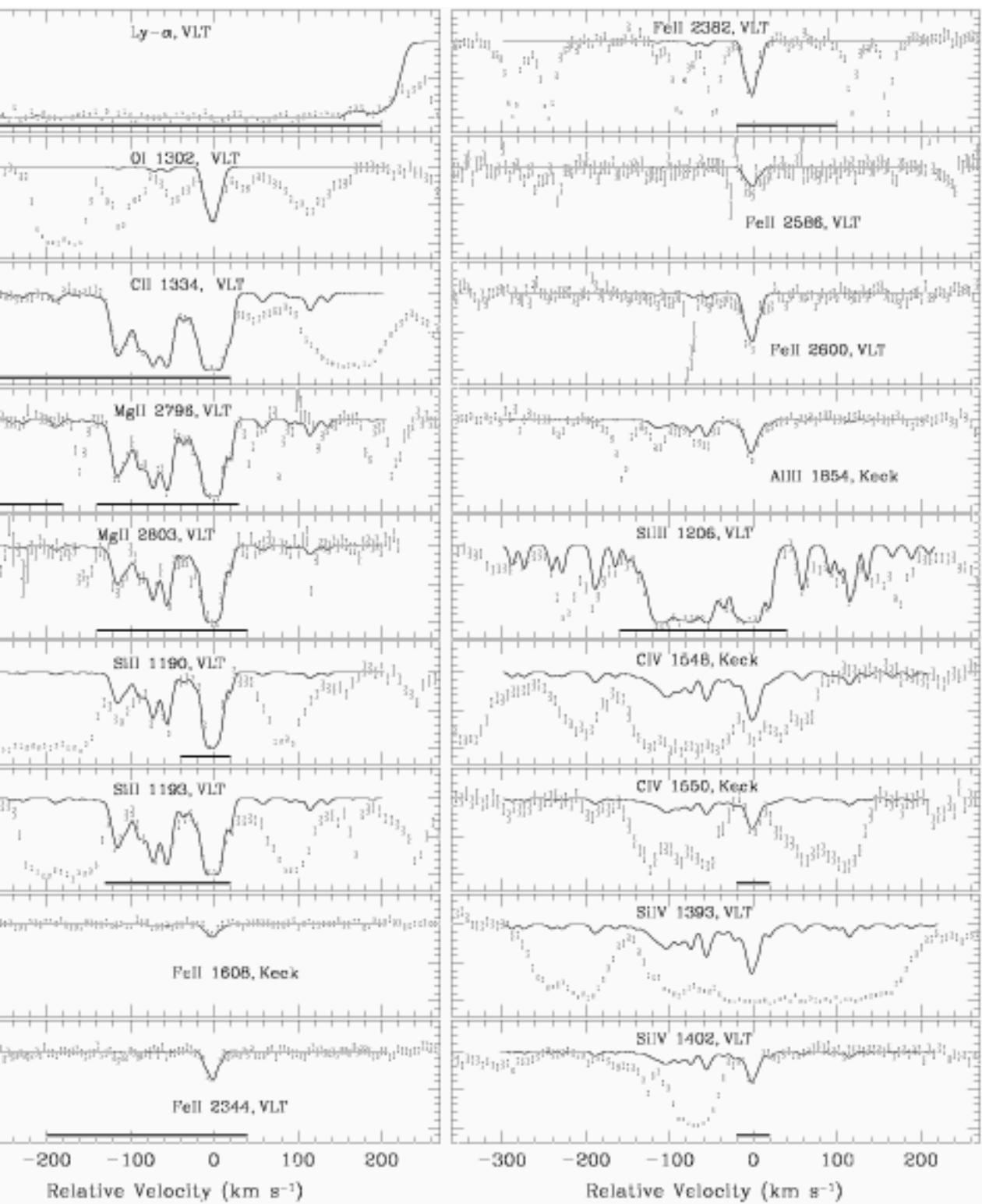
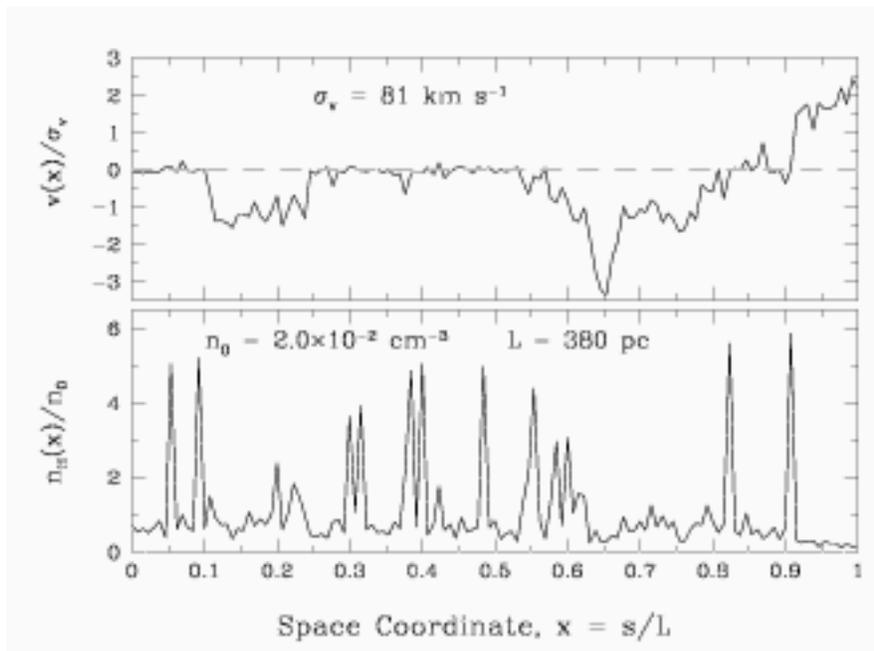
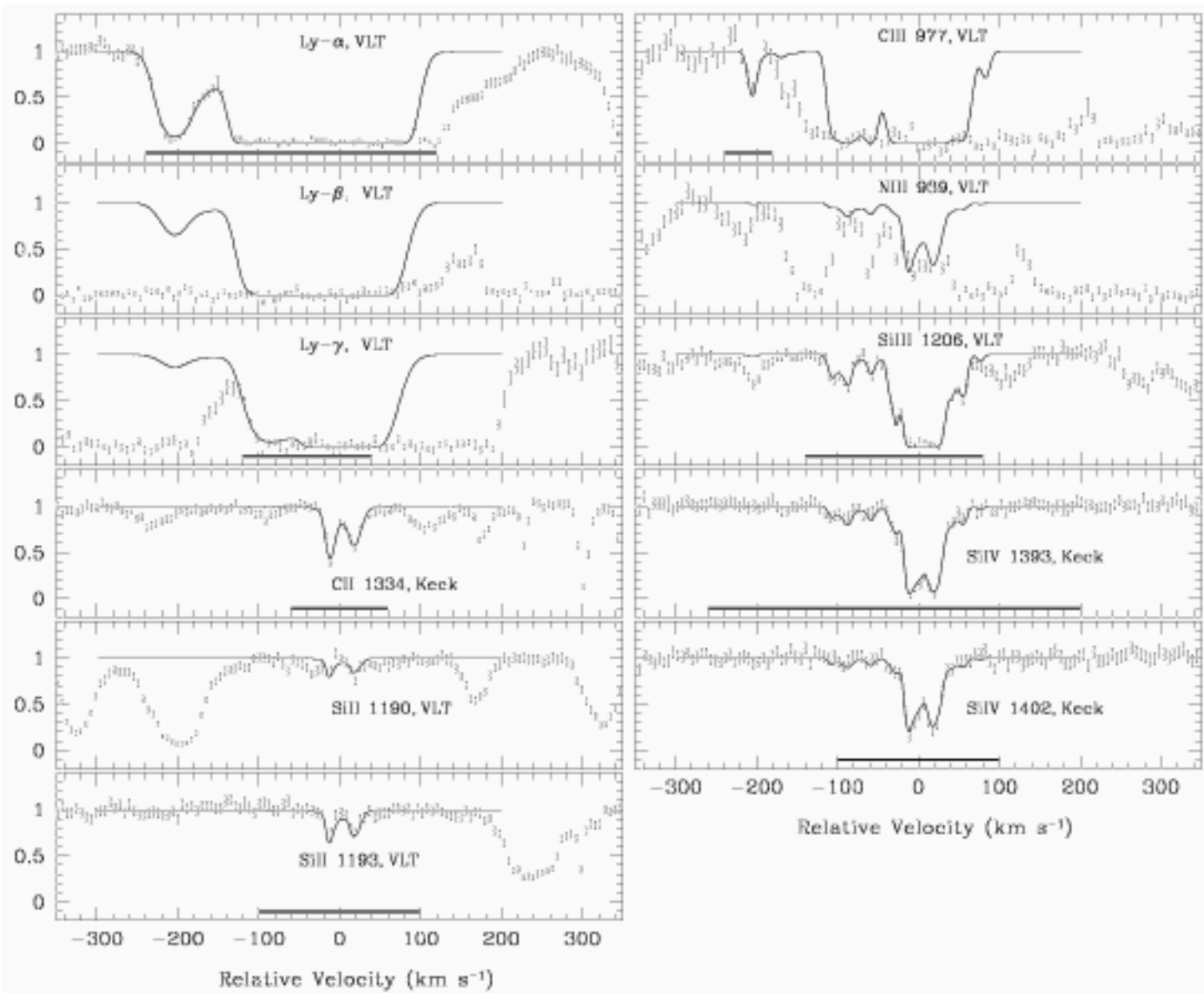
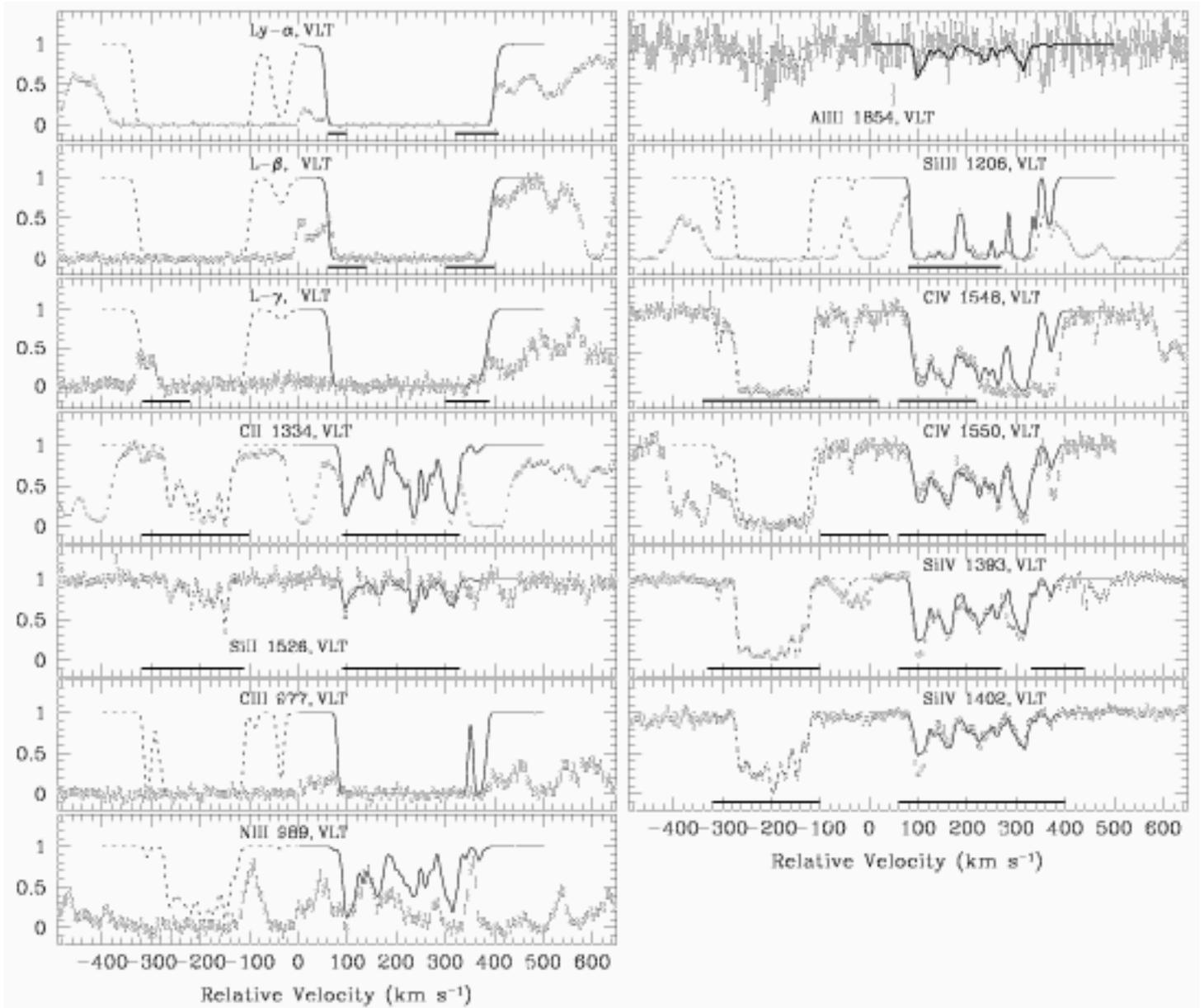
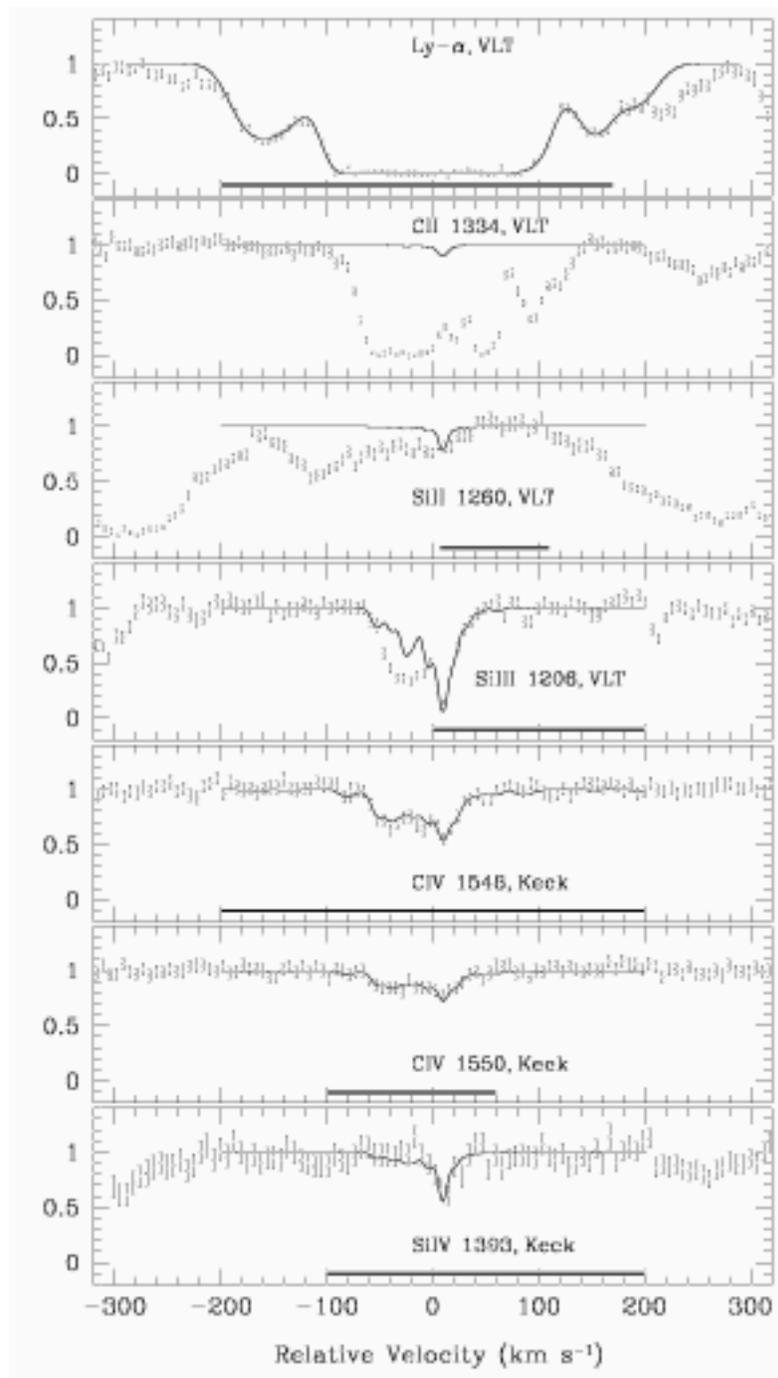
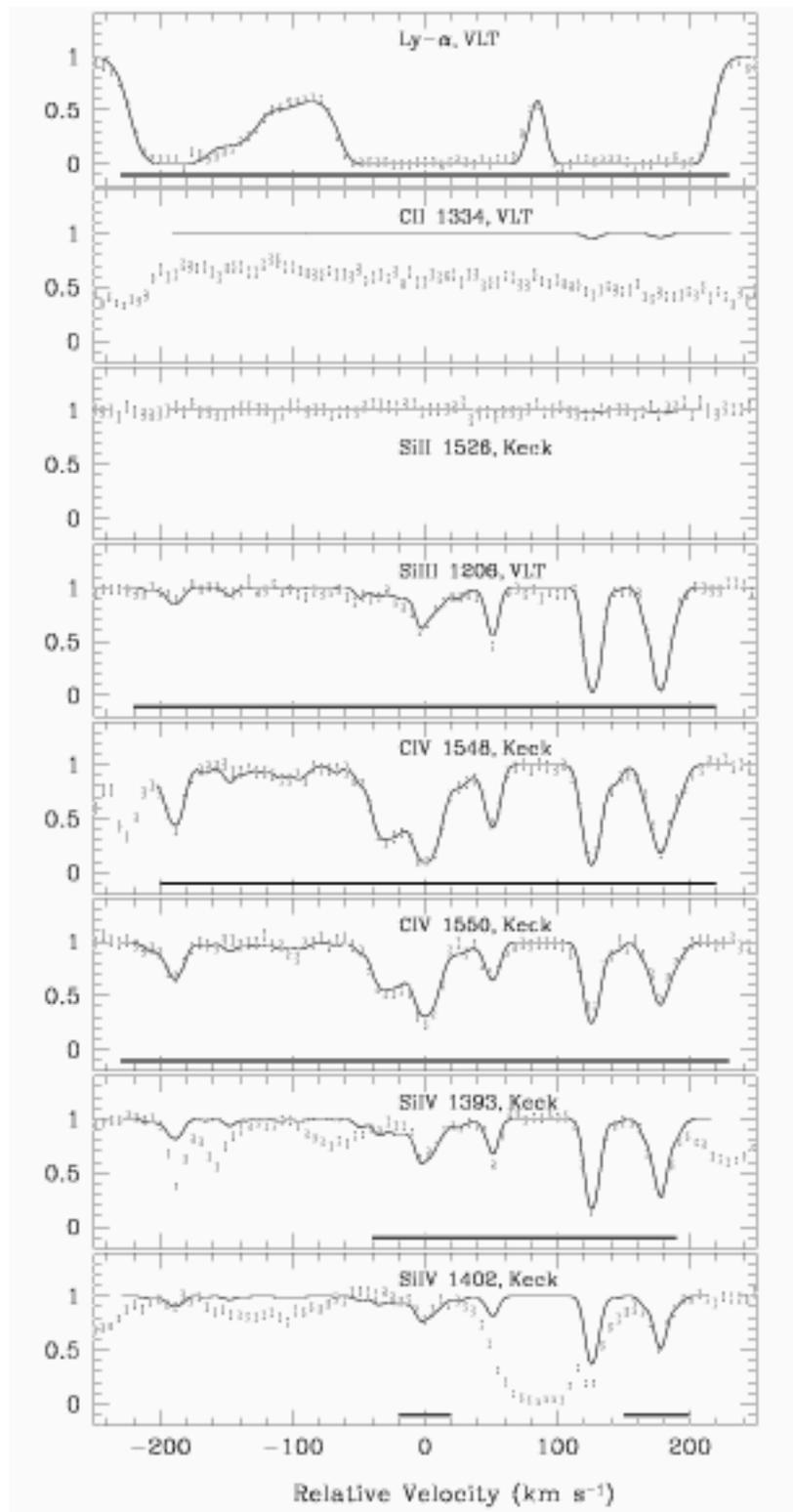
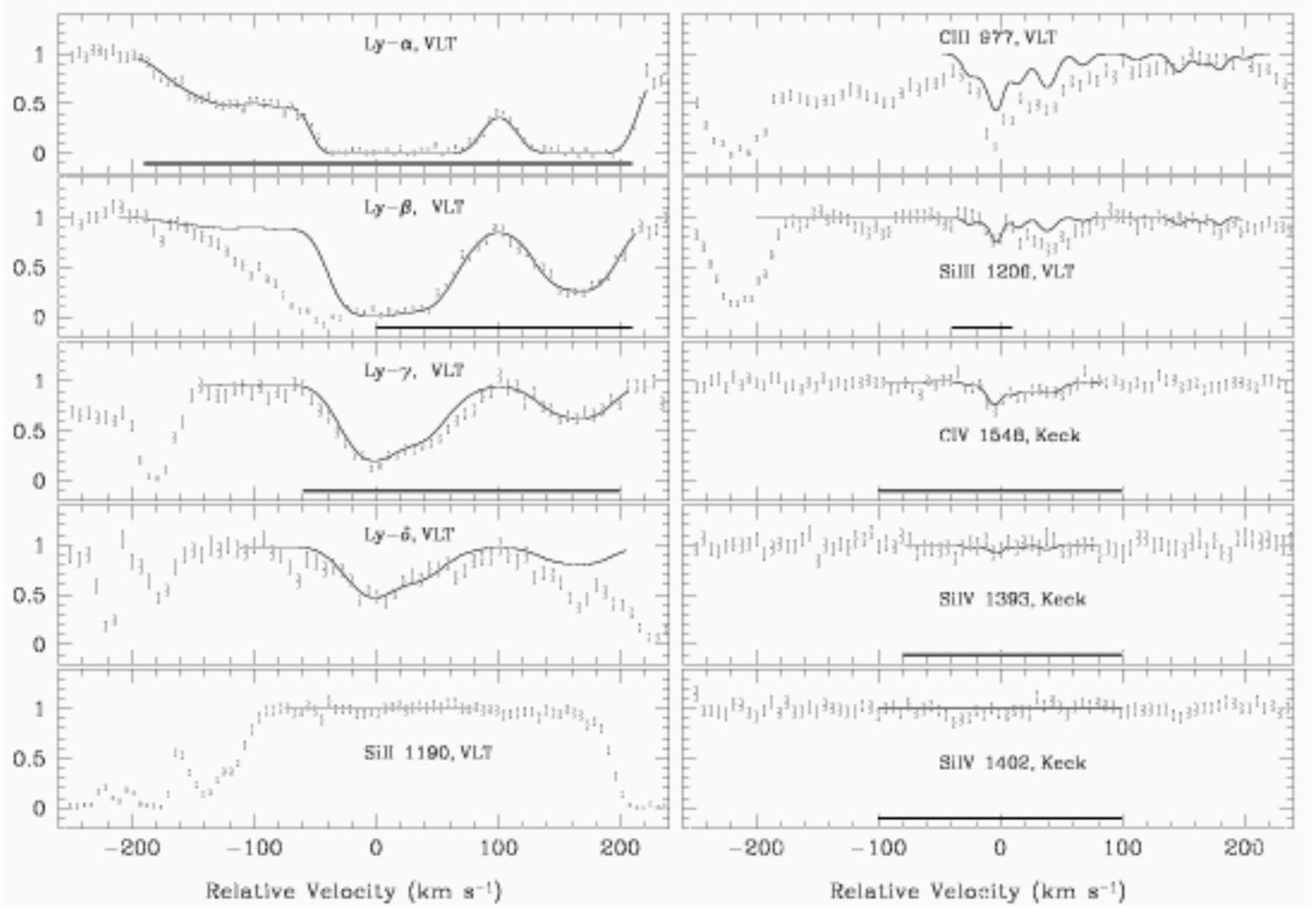
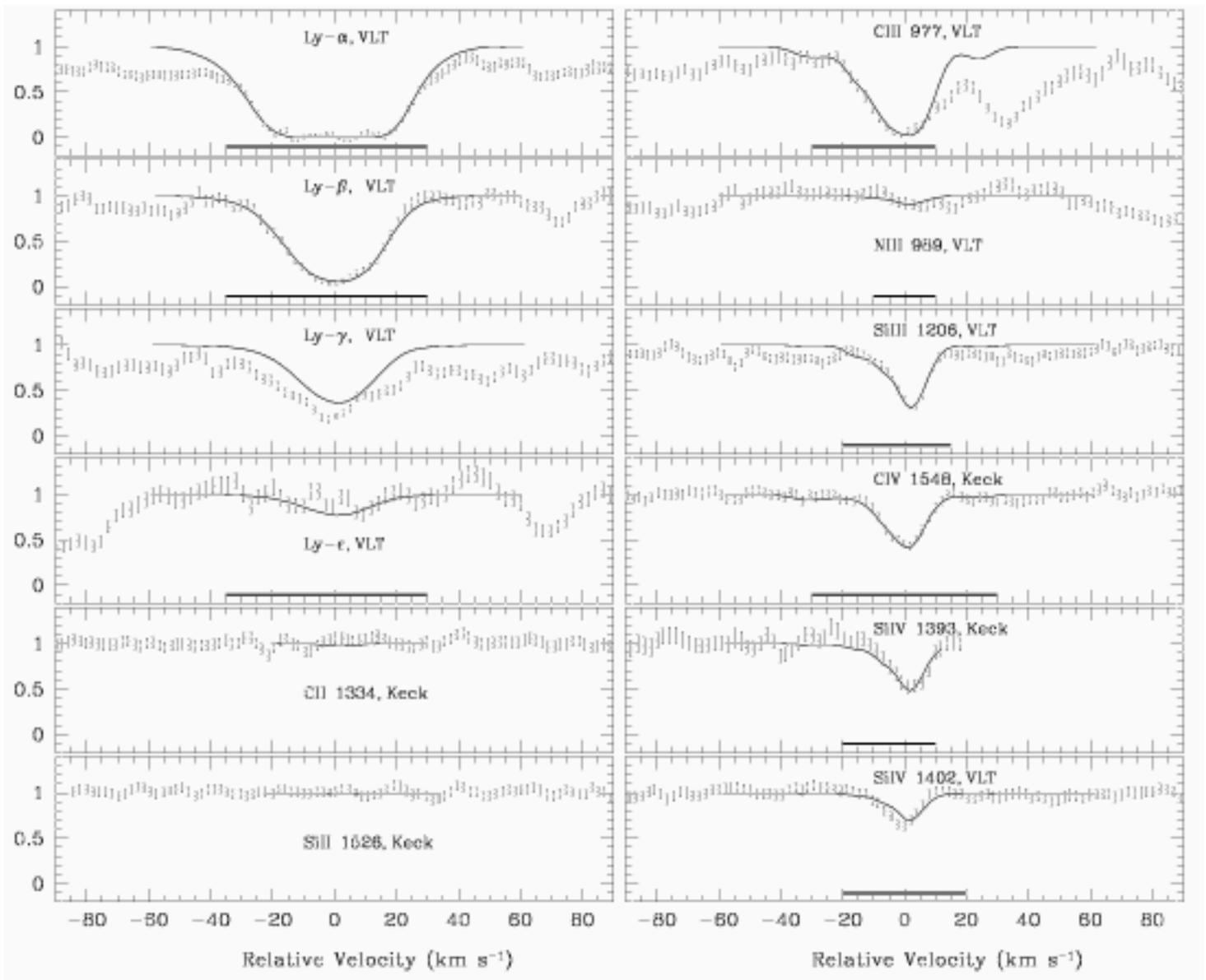
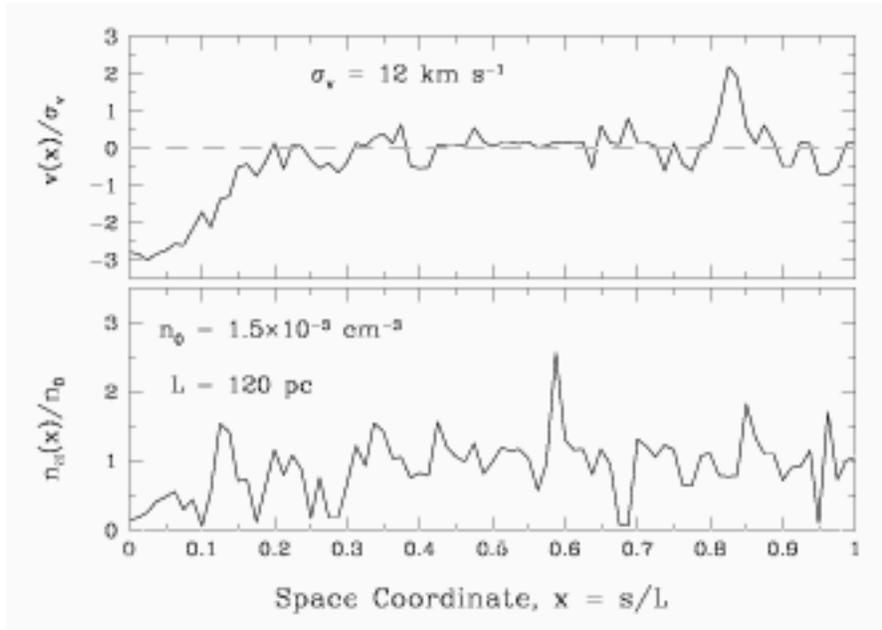
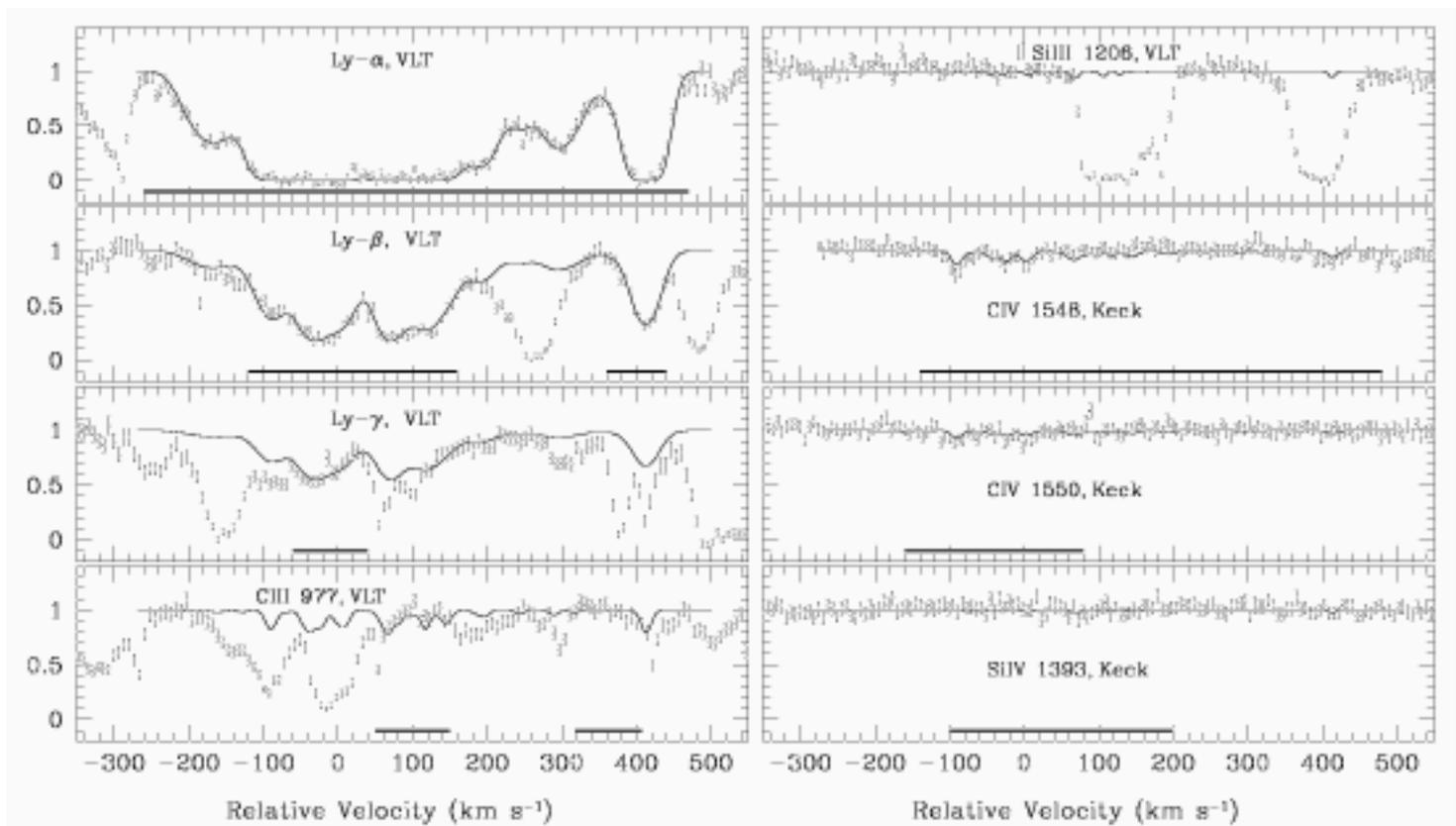
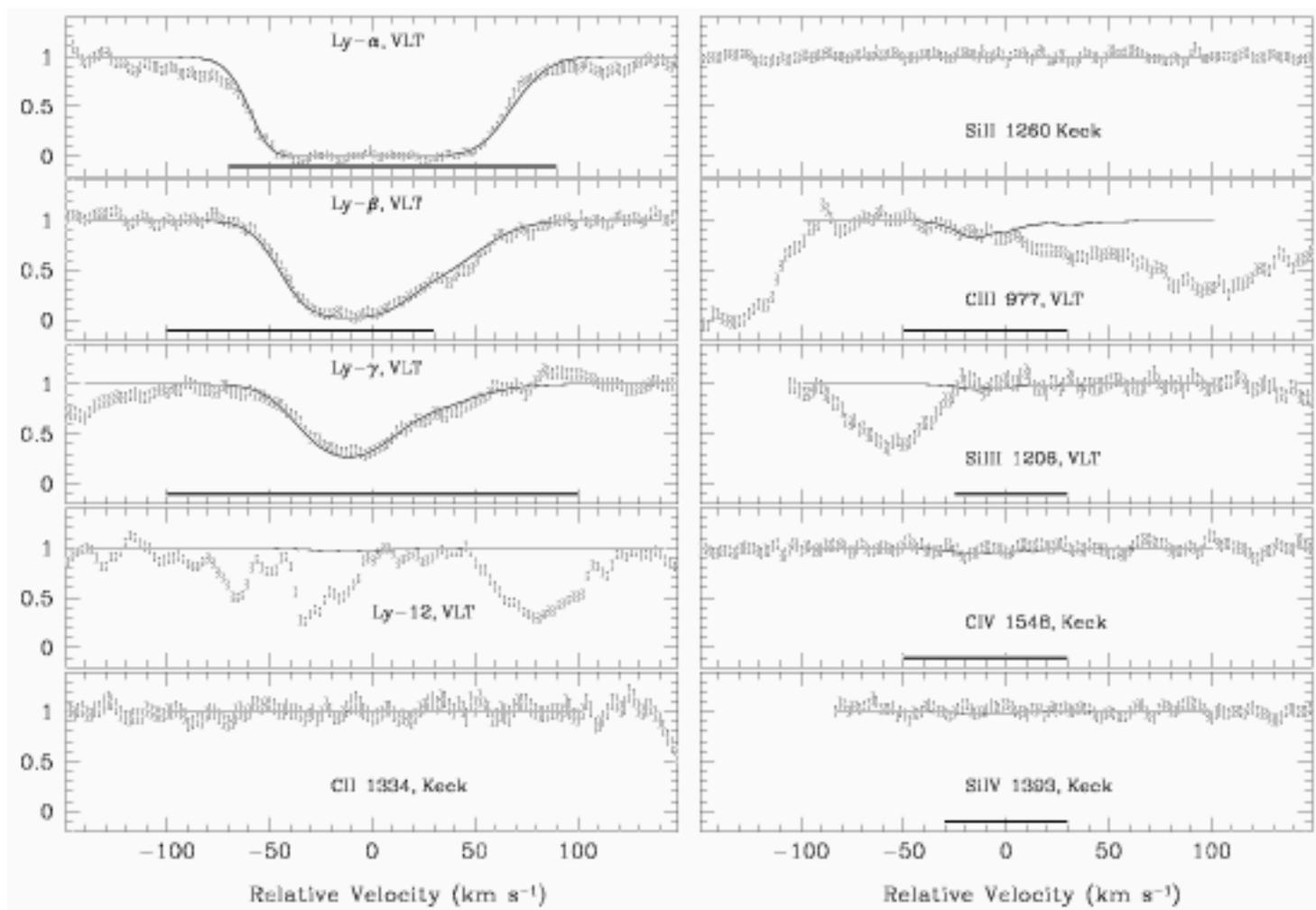
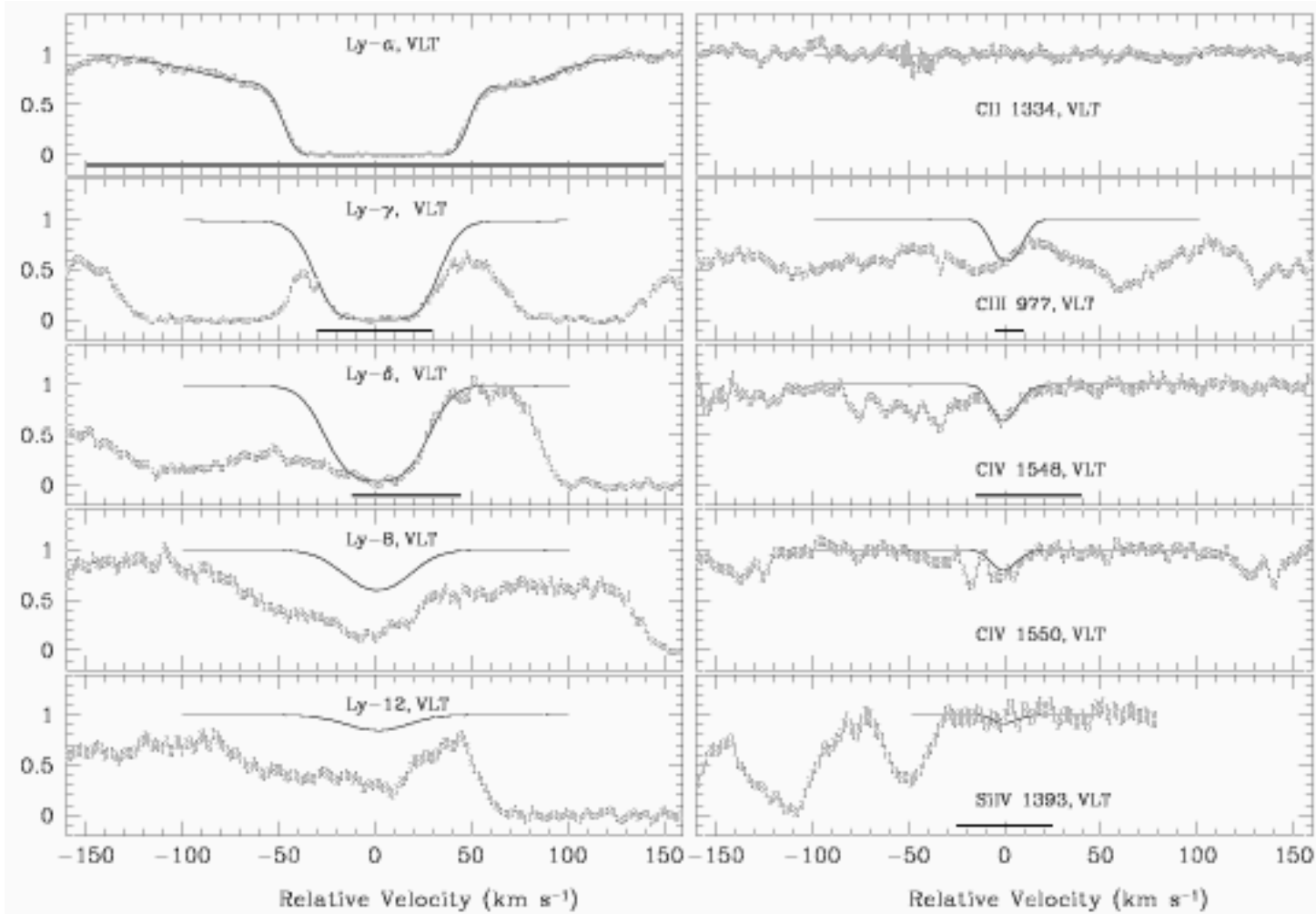
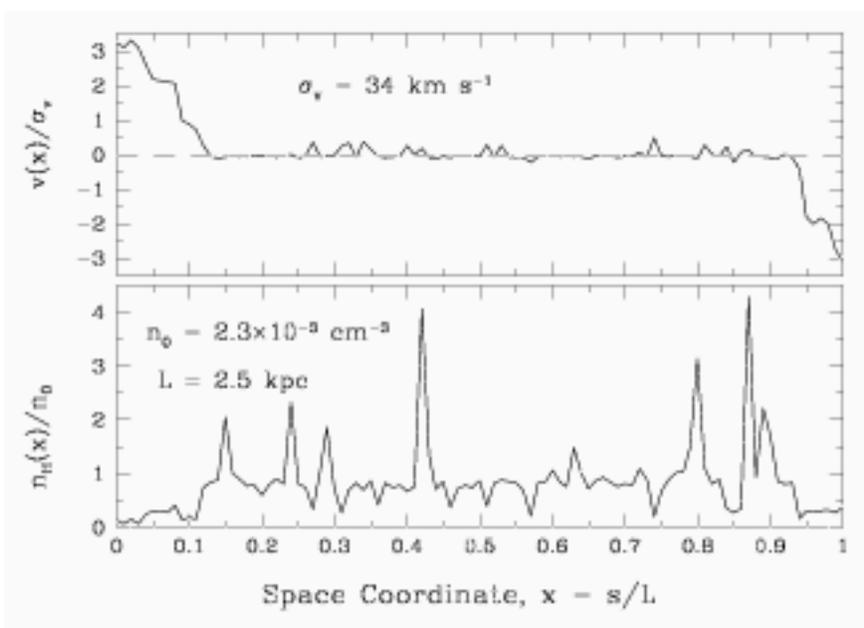
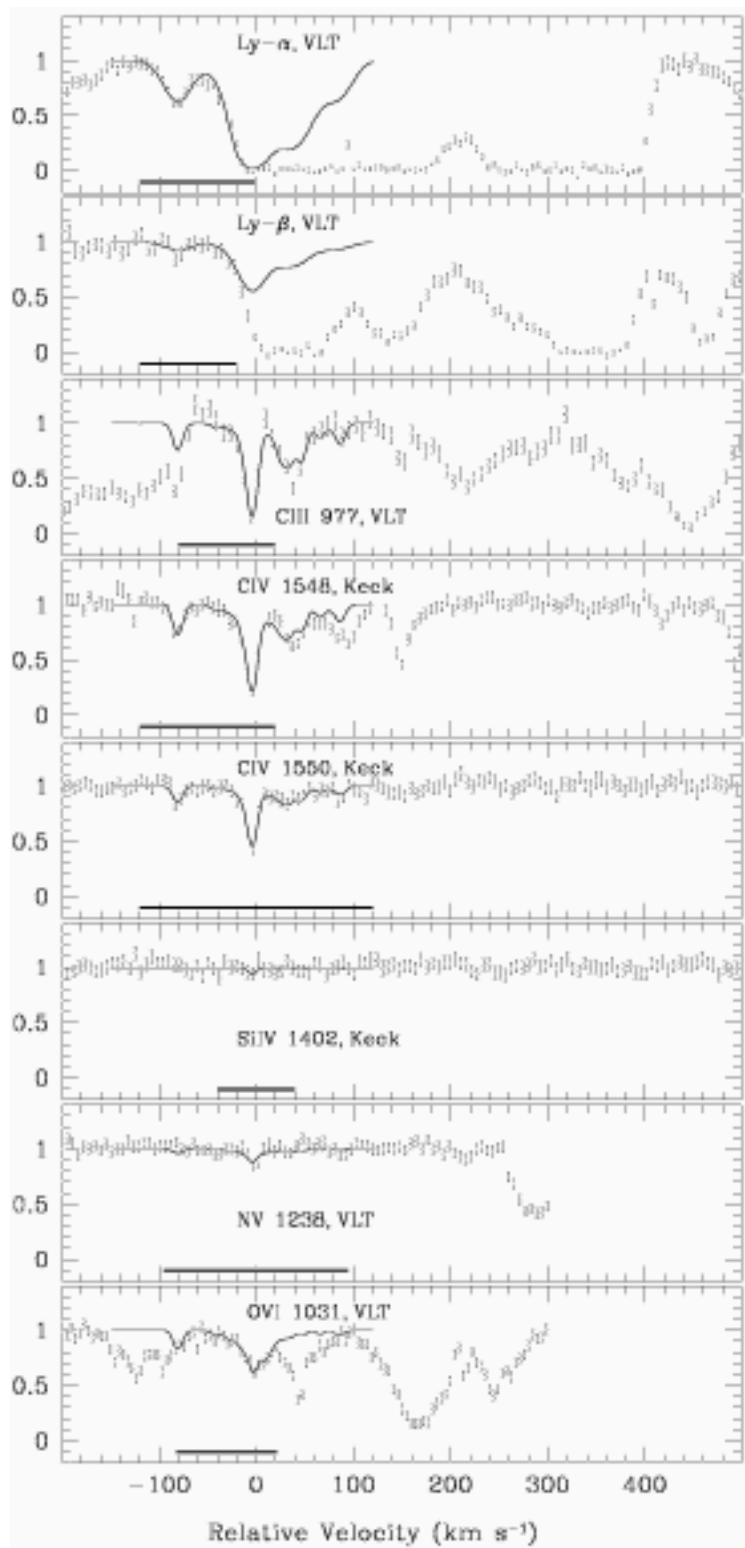
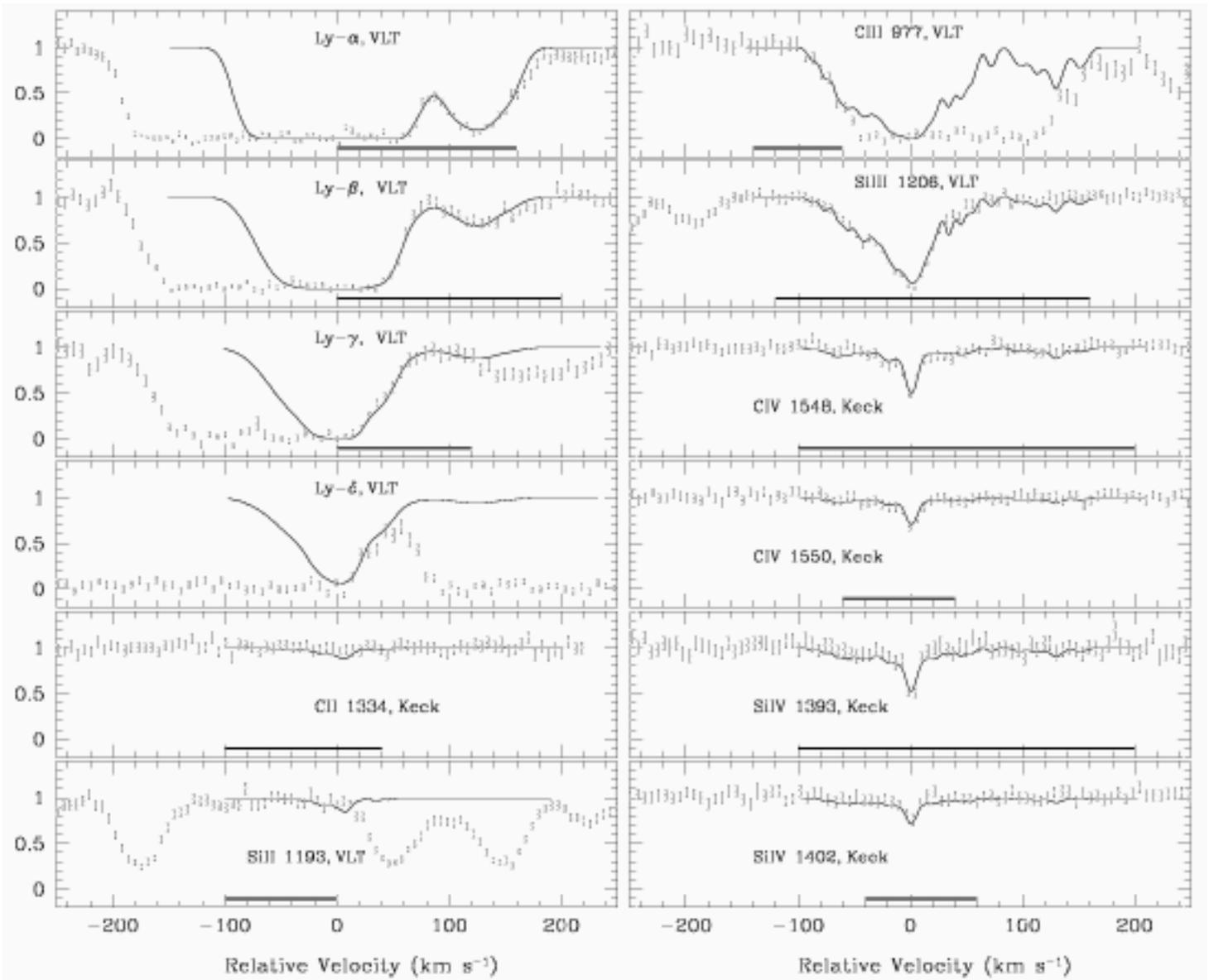
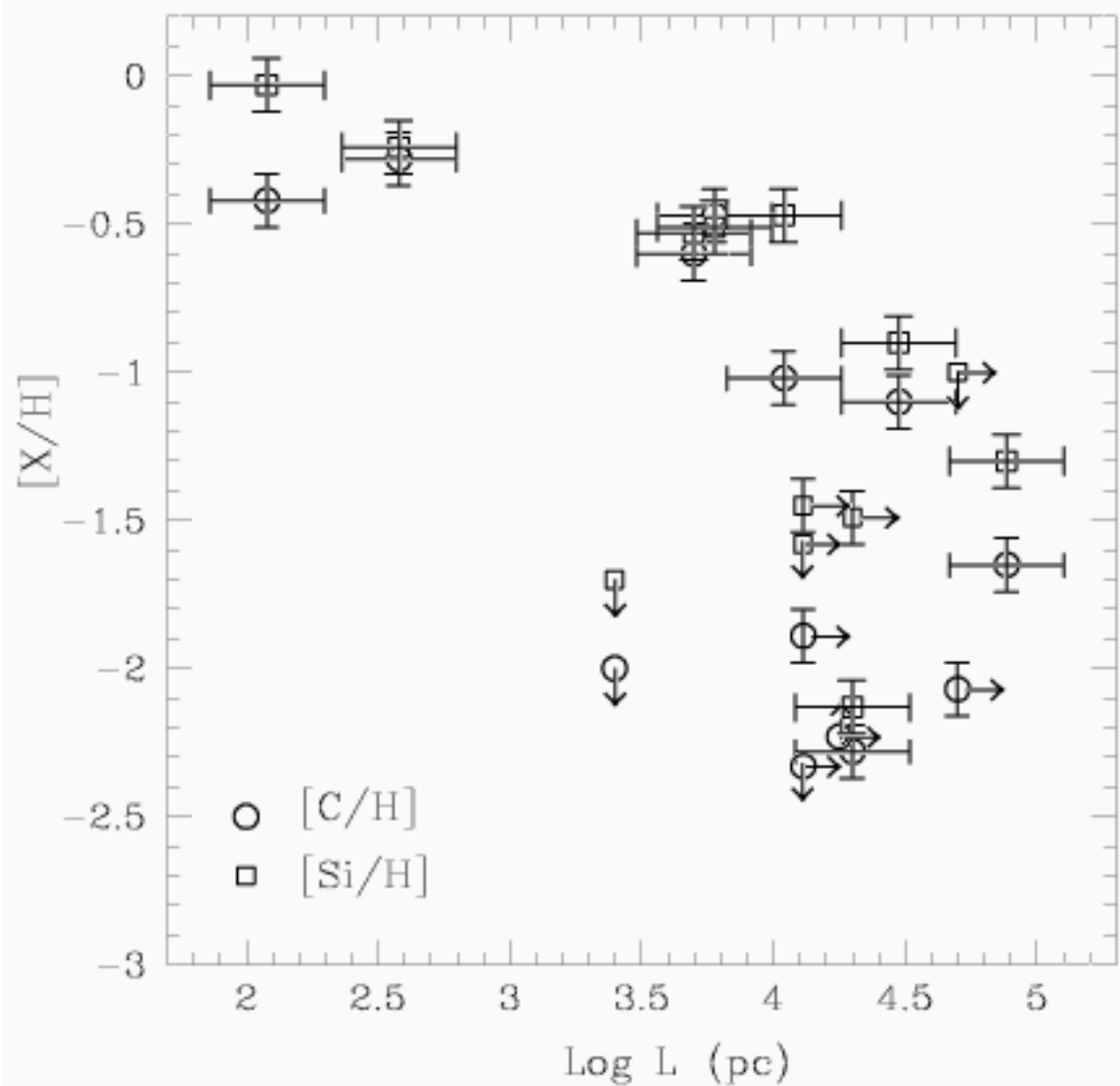
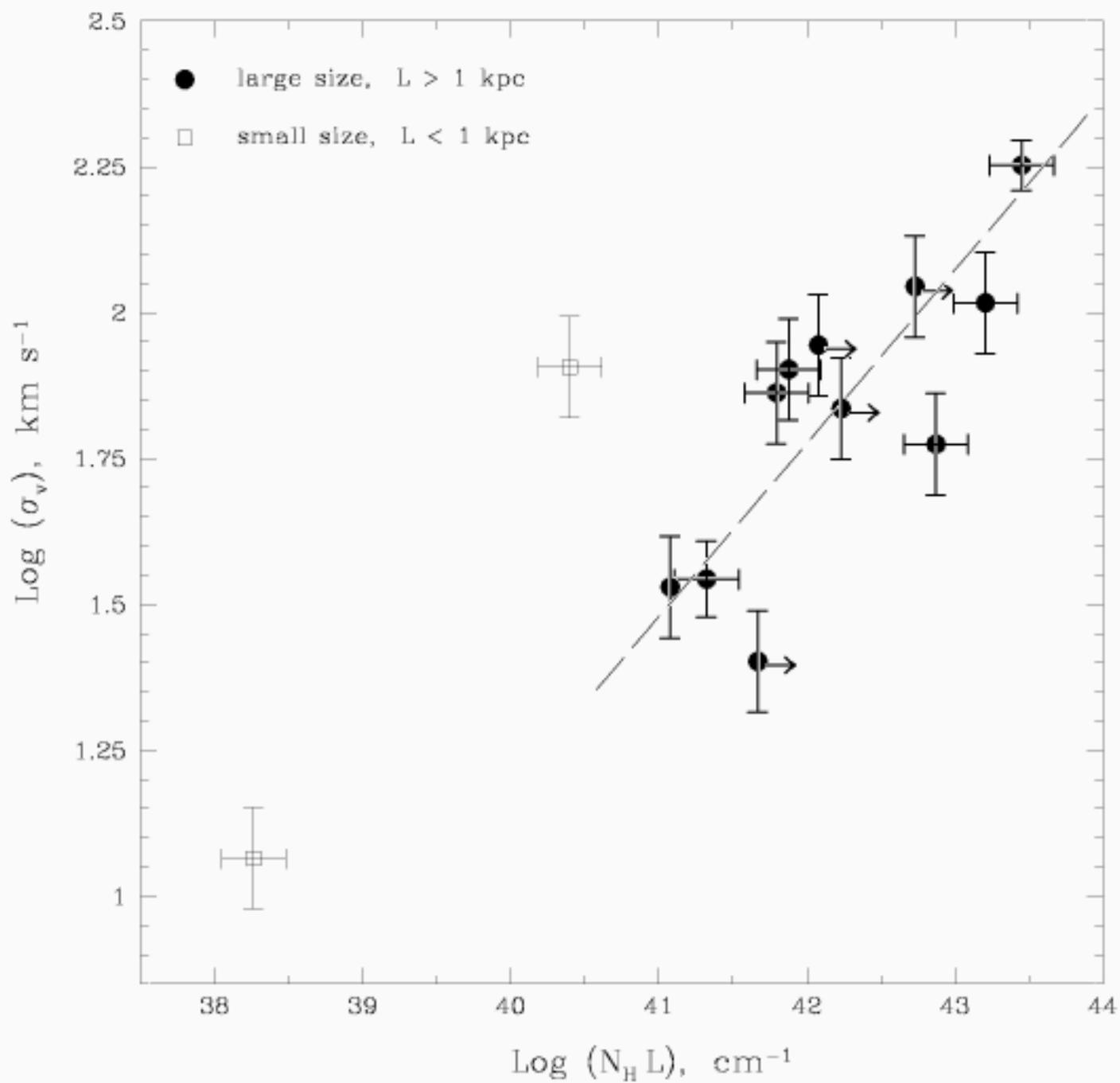
\end{document}